%% file: main3.tex
\PassOptionsToPackage{hyphens}{url}
\documentclass[sigconf,screen,nonacm]{acmart}

\usepackage{amsmath}

\usepackage{algorithm}
\usepackage{algorithmic}
\renewcommand{\algorithmicrequire}{\textbf{Input:}}
\renewcommand{\algorithmicensure}{\textbf{Output:}}
\usepackage{graphicx}
\usepackage{textcomp}
\usepackage{xcolor}
\usepackage[normalem]{ulem}
\usepackage{booktabs}
\usepackage{float}
\usepackage{array}
\usepackage{enumitem}
\usepackage{tikz}
\usepackage{comment}
\usetikzlibrary{arrows.meta,patterns,positioning,decorations.pathreplacing,backgrounds,fit,calc}
\input{fig_styles}
\usepackage{soul}
\setstcolor{red} 

\usepackage{placeins}      % \FloatBarrier keeps floats from drifting past the references

\makeatletter
\newcommand{\res}[1]{\@ifundefined{res@#1}{\textbf{??#1??}\PackageWarning{res}{undefined key #1}}{\@nameuse{res@#1}}}
\makeatother

\expandafter\def\csname res@c/band/islfactormin\endcsname{3.2} 
\expandafter\def\csname res@c/band/islfactormax\endcsname{8.2}
\expandafter\def\csname res@c/band/islshort\endcsname{1{,}024} 
\expandafter\def\csname res@c/band/isllongmin\endcsname{32{,}768} 
\expandafter\def\csname res@c/band/isllongmax\endcsname{65{,}536} 
\expandafter\def\csname res@c/band/oslfactorallislmin\endcsname{6.81} 
\expandafter\def\csname res@c/band/oslfactorallislmax\endcsname{12.2}

\begin{document}

\newcommand{\iscasubmissionnumber}{1782}

\pagenumbering{arabic}

\title{Analytical Modeling and Optimal GPU Provisioning for Disaggregated LLM Inference}

% \author{ISCA 2027 Submission \textbf{\#\iscasubmissionnumber} -- Confidential Draft -- Do NOT Distribute!!}
\author{Akshat Pandey}
\affiliation{%
  \institution{Indian Institute of Technology Bombay}
  \city{Mumbai}
  \country{India}
}

\author{Dominic Cheng}
\affiliation{%
  \institution{NVIDIA}
  \country{Santa Clara, USA}
}

\author{Sidharth Sharma}
\affiliation{%
  \institution{Indian Institute of Technology Jodhpur}
  \city{Jodhpur}
  \country{India}
}

\author{Ashwin Gumaste}
\affiliation{%
  \institution{NVIDIA}
  \country{Santa Clara, USA}
}

\begin{abstract}
\textbf{We present OSplit, an exact analytical method that enables efficient partitioning of a GPU L1 domain, into prefill and decode workers, including sharding for classical and advance GPUs. OSplit takes into consideration model parameters enabling any inference model to be partitioned into prefill and decode workers with a goal to produce the maximum throughput possible. OSplit is based on a capacity model that takes as inputs the model, the loading scheme (batch size, max sequence length) and models an inference serving framework (vLLM) and computes the efficiency of each GPU. For this, we model H100 and GB300 GPUs, and model a scale up domain, along with flash attention based kernels. OSplit first computes an optimal collective generation policy, which gives inputs to the capacity model that enables optimal split of prefill and decode GPUs. We evaluate our scheme over H100 (NVL8) and GB300 (NVL72) using vLLM inference serving and show throughput, predicted and measured latency showing near perfect theoretical-to-measured values over leading LLMs (Llama and Qwen).}
\end{abstract}

\maketitle

\section{Introduction}
\label{sec:intro}

Disaggregated inference has emerged as an effective approach to serving large language models (LLMs)~\cite{splitwise,beyondbuzz,sara}. An LLM request passes through two distinct phases: \emph{prefill}, which processes the input prompt and constructs key-value (KV) cache, and \emph{decode} that generates autoregressive response. 
Prefill is compute-intensive, whereas decode is constrained by memory and bandwidth (BW)~\cite{splitwise}. Disaggregated serving places these phases on separate GPU worker \textit{pools}, allowing each to be optimized for high throughput while satisfying latency constraints. 
Disaggregation enables heterogeneous deployments, allowing CSPs to use state-of-the-art GPUs for prefill and older GPUs or specialized accelerators, such as LPUs, for decode.
Such specialization offers opportunities to improve time to first token (TTFT), time between tokens (TBT), GPU utilization, and ultimately tokens per second per dollar, while reducing capital and operating expenditure. 

Realizing these advantages, however, requires answering fundamental system-design questions. \emph{First, how should a fixed GPU budget be divided between prefill and decode?} This allocation must be efficient within a rack or a larger scale-up domain and must reflect the different processing demands of the two phases. \emph{Second, how should the GPUs assigned to each phase be organized into model instances and parallelism groups?} 
Tensor, data, and, for mixture-of-experts (MoE) models, expert parallelism (EP) affect the amount of computation, model-weight storage, communication, and request concurrency each pool can sustain. 

The allocation problem arises because every request must traverse both phases. In disaggregated serving, the prefill pool processes the prompt and transfers the resulting KV cache to a decode pool, which generates output tokens. Consequently, the steady-state request throughput of the deployment is limited by the pool with lower processing capacity. Adding GPUs to a pool that already has excess capacity does not improve end-to-end throughput. Therefore, assigning those GPUs to the bottleneck pool can instead increase throughput until the two capacities are balanced. %Thus, choosing the prefill--decode split is not simply a matter of dividing GPUs in proportion to the number of tokens processed by each phase.
The best split is not fixed and depends on model architecture, parameters, GPU characteristics, serving framework, input and output sequence lengths (ISL, OSL). The parallelism configuration of each pool further changes its effective capacity. Although balancing the two pools suggests an ideal allocation, GPUs must in practice be assigned to complete model instances with feasible tensor, data, and expert parallel configurations. 
The resulting optimization is therefore coupled and discrete: a seemingly balanced GPU count may be infeasible or may perform worse than another allocation once memory requirements, communication costs, and serving constraints are considered. This matters especially for an AI factory serving heterogeneous agentic workloads, where prompt and response lengths can vary sizably~\cite{chang2026llm}.

An operator may therefore maintain several disaggregated deployments, each configured for a range of request lengths, and route incoming prompts to an appropriate deployment. Historical request-length histograms can inform the choice of these ranges, while the resource allocation within each deployment must be determined from its workload and system characteristics. Existing approaches typically profile the model on target hardware, fit performance models to measurements, and search candidate configurations~\cite{splitwise,distserve,vidur,aiconfigurator}. Such measurements are computationally expensive and must be revisited when the model, GPU, serving framework, or workload changes. %This limits the ability to explore design choices or rapidly reconfigure deployments.

We present \emph{OSplit}, an analytical approach to GPU partitioning and parallelism selection for disaggregated inference. Given a GPU budget, workload characteristics, model specifications, and hardware and serving-framework parameters, OSplit estimates the processing capacity of the prefill and decode pools and selects a configuration that maximizes the lower of their throughputs. Rather than relying on an exhaustive set of deployment measurements, the approach uses an analytical capacity model, together with the required hardware parameters, to reason about the computational, memory, and communication costs of candidate configurations.

OSplit enumerates feasible parallelism configurations for each phase, accounting for model compatibility, GPU memory capacity, and optional TBT constraints. Using the capacity model developed in Section~\ref{sec:model}, it estimates the GPU time required per prefill and per decoded output token. These costs reveal the ideal continuous allocation that balances the two pools. Because this allocation may not correspond to an integer number of complete model instances, OSplit subsequently examines feasible combinations of prefill and decode parallelism and their realizable GPU allocations, selecting the combination that maximizes the minimum phase throughput. Its output comprises the number of GPUs assigned to each pool, the corresponding parallelism degrees, and inference throughput.

Our primary deployment target is an L1 scale-up domain, such as a GPU rack, although the analytical formulation is intended to extend to larger domain sizes. The approach provides a way to reason about provisioning before executing every candidate configuration on the target deployment. It also makes explicit why a preferred split changes with the workload, model, and hardware, rather than treating the configuration as the output of a black-box search. We validate the model using vLLM on H100 GPUs and a GB300 NVL72,  and study how the selected split varies with workload, model, and L1-domain size.

The paper is organized as follows. \S~\ref{sec:problem} describes the system architecture, execution model, and provisioning problem. \S~\ref{sec:model} derives the capacities of both pools, and \S~\ref{sec:prov} develops OSplit. \S~\ref{sec:impl} describes its implementation and vLLM integration; \S~\ref{sec:eval} validates the model on H100 GPUs and GB300 NVL72; and \S~\ref{sec:results} examines how workload, model, and L1-domain size affect the selected split. \S~\ref{sec:related} discusses related work, and \S~\ref{sec:conc} concludes the paper.

\section{System Architecture and Problem Formulation}
\label{sec:problem}

This section describes the GPU deployment setting and the organization of prefill and decode workers. We then introduce the supported parallelism configurations and formulate the problem of allocating a fixed GPU budget to maximize inference throughput under memory and latency constraints.

\subsection{Deployment Setting and Communication Domain}

We consider an AI factory comprising GPU compute nodes interconnected through a high-bandwidth scale-up fabric. Each compute node contains CPUs and GPUs, with NVLink enabling GPU-to-GPU communication. Rack-scale systems may organize GPUs and NVLink switch ASICs in a fat-tree topology. We define a \emph{GPU domain} as the fixed set of $N$ GPUs provisioned by OSplit. Our evaluation considers two configurations: an eight-GPU H100 node NVL8 and a 72-GPU GB300 NVL72 rack. OSplit determines the allocation of GPUs between prefill and decode workers within a domain, while the hardware topology and interconnect bandwidth are fixed inputs to the performance model.

\subsection{Request execution and model instances}

External requests enter through the data-center network and are routed to serving hosts, where vLLM~\cite{vllm} organizes them into batches and launches the GPU kernels. Each request undergoes prefill, which processes the input prompt and generates the KV cache, followed by autoregressive decoding. The workload is characterized by the mean ISL $I$ and mean OSL $\bar{O}$.

A GPU domain may host one or more model instances, depending on the model size, precision, and available GPU memory. Multiple instances enable concurrent processing of independent batches. Although some implementations favor power-of-two parallelism degrees, OSplit does not impose this restriction in its general formulation; deployment-specific feasibility constraints may nevertheless restrict the choices.

% We assume vLLM~\cite{vllm} executes on the CPU hosts in the compute nodes and enables firing of model kernels to corresponding GPUs. %In disaggregated serving, a prefill instance processes a batch of prompts and transfers each request's KV cache to a decode instance, which generates the remaining output tokens. 
% Prefill performs a forward pass over the prompt and produces the first token; decode performs successive iterations, each generating one token per active request, for example using top-$p$ or top-$k$ sampling till an EOS token is generated. Requests with similar input lengths may be grouped into batches; the workload assigned to a GPU domain is described by a distribution of prompt lengths with mean $I$ and output lengths with mean $\bar O$.

% An LLM may occupy the entire GPU domain, or several model instances may coexist when its memory footprint permits. Modern large dense and MoE models, including DeepSeek, Kimi, Llama, and GPT-family architectures, motivate both possibilities, depending on model size, precision, and GPU memory. \ash{add ref for these models}
% For example, a 72-GPU domain with 384 GB of HBM per GPU may accommodate multiple instances for some model configurations. Multiple instances enable concurrent processing of different batches. Some implementations favor power-of-two parallelism groups, but OSplit does not impose this restriction in its general formulation; deployment-specific feasibility constraints may nevertheless limit the choices.

\subsection{Tensor, data, and expert parallelism}

Within a model instance, tensor parallelism (TP) partitions layer weights and attention computations across $g_t$ GPUs, with each GPU performing a portion of the operations~\cite{megatronlm}. Data parallelism (DP) replicates computation across $g_d$ groups so that different batches can be processed concurrently. For dense models, our terminology treats each independently replicated model as a separate serving instance: a dense instance has $g_d=1$ and occupies $G=g_t$ GPUs. Deploying additional dense-model copies therefore increases the number of instances, rather than the DP degree within an instance.

For mixture-of-experts (MoE) models, EP distributes experts of a layer across GPUs~\cite{gshard}. In the configuration modeled, $G=g_tg_d$ GPUs of an MoE instance form one EP group. Attention is replicated over $g_d$ groups, each TP-ed over $g_t$ GPUs, while experts are distributed across the full $G$-GPU group. Thus the $g_d$ attention replicas within an MoE instance are not equivalent to $g_d$ independent dense-model instances: they share an EP organization. The instance can be viewed as a $g_t\times g_d$ grid for attention and one EP group for the experts. The model's expert load balancer, captured later by $\Lambda$, affects its performance. OSplit chooses TP/DP configurations for each phase, with the corresponding EP organization determined by this model.

Each prefill instance uses $G_{\mathrm{PF}}=g_t^{\mathrm{PF}}g_d^{\mathrm{PF}}$ GPUs, and each decode instance uses $G_{\mathrm{DC}}=g_t^{\mathrm{DC}}g_d^{\mathrm{DC}}$ GPUs. Within each pool, we assume regular instances have the same parallelism configuration. A deployment may also bin-pack a smaller, lower-priority best-effort instance into leftover GPUs. 
%For example, four 16-GPU instances and one 8-GPU instance fill a 72-GPU domain. This heterogeneous-instance packing is a practical deployment option rather than a decision represented by the regular-instance optimization below.

As an illustration, suppose OSplit assigns 24 of 72 GPUs to prefill and 48 to decode. Under a deployment restriction to power-of-two TP groups and applicable latency constraints, one feasible organization is three 8-GPU prefill instances and three 16-GPU decode instances, thus adding up to 72 GPUs. This example demonstrates how a pool-level allocation maps to complete instances.

\begin{table*}[t]
  \centering
  \small
  \caption{Notations. A subscript or superscript PF or DC refers to the prefill or the decode pool.}
  \label{tab:notation}
  \vspace{-10pt}
  \begin{tabular}{@{}l >{\raggedright\arraybackslash}p{0.36\textwidth} @{\qquad}l >{\raggedright\arraybackslash}p{0.36\textwidth}@{}}
    \toprule
    \multicolumn{2}{@{}l}{\emph{Workload}} & \multicolumn{2}{l}{\emph{LLM}} \\
    $I$ & mean number of tokens in a prompt & $L$ & number of layers \\
    $\bar O$ & mean number of tokens in a response & $H_q$, $H_{kv}$ & query heads; KV heads \\
    $S$ & mean length of a request's KV cache during decode (in tokens)& $h_{kv}$ & KV heads per GPU\\
    \multicolumn{2}{@{}l}{\emph{L1 domain and parallelism}} & $d_h$ & size of one attention head \\
    $N$; $m$, $n$ & number of GPUs provisioned (normally one L1 domain); number of prefill GPUs, number of decode GPUs& $E$; $k$ & experts per layer; experts each token is sent to \\
    $g_t$, $g_d$ & tensor and data parallel degree of an instance& $W$ & model weight bytes per GPU\\
    $G$ & GPUs per instance = $g_t g_d$& $\kappa$ & KV-cache bytes per token per GPU\\
    \multicolumn{2}{@{}l}{\emph{Serving framework}} & \multicolumn{2}{l}{\emph{GPU}} \\
    $T_b$ & token budget of one prefill iteration & $M_{\mathrm{HBM}}$ & HBM memory capacity\\
    $U$ & fraction of GPU memory the framework may use & $F$; $N_{\mathrm{SM}}$ & peak FP8 tensor-core throughput; number of SMs \\
    $M_{\mathrm{res}}$ & memory the framework reserves for itself & $\beta_H$, $\beta_{L2}$ & achieved HBM and L2 bandwidth \\
    \midrule
    \multicolumn{2}{@{}l}{\emph{Batches and times}} & \multicolumn{2}{l}{\emph{Capacities and provisioning}} \\
    $B$ & batch size& $c_{\mathrm{PF}}$, $c_{\mathrm{DC}}$ & requests per second one GPU of a pool sustains \\
    $b$ & batch size per copy of the model = $B/g_d$& $\phi$ & throughput of the L1 domain, requests per second \\
    $R$ & token rows a layer processes & $r^*$ & best ratio of prefill to decode GPUs \\
    $\tau_{\mathrm{PF}}$ & prefill iteration time (time of one forward pass over a batch of prompts) & $P$ & prefill GPU-seconds per request \\
    $\tau_{\mathrm{DC}}$ & decode iteration time (time of one forward pass that produces one token per request) & $D$ & decode GPU-seconds per output token \\
    $\tau$ & limit on the time between tokens & $\Lambda$ & expert load of the busiest GPU relative to the mean \\
    \bottomrule
  \end{tabular}
   \vspace{-10pt}
\end{table*}

\subsection{Provisioning problem}

Because every request passes through prefill and decode, the sustainable request rate is the minimum of the two pool capacities. For a given workload and parallelism choice, let $c_{\mathrm{PF}}$ and $c_{\mathrm{DC}}$ denote the sustainable request rates per allocated GPU in the respective pools. These quantities depend on the phase-specific TP/DP configurations and the analytical cost model developed in \S~\ref{sec:model}. If there are $k_{\mathrm{PF}}$ prefill instances and $k_{\mathrm{DC}}$ decode instances, their GPU allocations are:
\begin{equation*}
    m = k_{\mathrm{PF}}G_{\mathrm{PF}}, \qquad
    n = k_{\mathrm{DC}}G_{\mathrm{DC}}.
\end{equation*}
We seek the split and parallelism configurations that maximize steady-state request throughput:
\begin{equation}
  \begin{aligned}
    \max\quad & \phi=\min\bigl(m\,c_{\mathrm{PF}},\; n\,c_{\mathrm{DC}}\bigr)\\
    \text{subject to}\quad & m+n\le N,\\
   % & m=k_{\mathrm{PF}}\,G_{\mathrm{PF}},\quad n=k_{\mathrm{DC}}\,G_{\mathrm{DC}},\\
    & \text{every instance fits in GPU memory},\\
    & \tau_{\mathrm{DC}}\le\tau.
  \end{aligned}
  \label{eq:problem}
\end{equation}
The optimization variables are the GPU allocations $m,n$, the four phase-specific parallelism degrees ($g_t^{\mathrm{PF}}$, $g_d^{\mathrm{PF}}$, $g_t^{\mathrm{DC}}$, $g_d^{\mathrm{DC}}$), and the integer instance counts. Here $\tau_{\mathrm{DC}}$ is the duration of a decode iteration and $\tau$ is the allowed TBT. 
The capacities are nonlinear functions of the parallelism configurations, making the problem a discrete, nonlinear provisioning problem. The formulation maximizes throughput under the stated constraints. 
The model uses tile-level GPU performance analysis to derive prefill and decode throughput from computation, memory access, and communication costs.
\S~\ref{sec:model} and \S~\ref{sec:prov} develop the OSplit procedure for obtaining the capacities and choosing a feasible allocation. Table~\ref{tab:notation} summarizes the notations used in the analysis. 

\subsection{Inputs, Outputs, and Assumptions}
\noindent\textbf{Inputs.} OSplit takes as input the LLM architecture and precision; GPU compute, memory, and interconnect characteristics; serving-framework memory reservations and prefill token budget; workload input/output length distributions; and the available GPU budget $N$ and domain topology.
We model vLLM~0.26.0~\cite{vllm026} and vLLM~0.32.0~\cite{vllm032} in \S~\ref{sec:impl}.\\
\input{fig_overview_1}
\noindent\textbf{Outputs.} OSplit returns the prefill-to-decode GPU ratio, the integer GPU allocation and parallelism configuration of each pool, and the predicted per-GPU and aggregate throughput.\\
\textbf{Assumptions and scope.}
We consider a single LLM deployed on homogeneous GPUs under steady-state, sufficiently high load, with a uniform parallelism configuration within each pool.
Hardware topology and interconnect characteristics are fixed inputs rather than optimization variables.
Workloads are characterized by their ISL/OSL distributions. OSplit can therefore evaluate deployment configurations without exhaustively profiling each candidate on the target hardware. %Figure~\ref{fig:overview} provides an overview of the OSplit method.
Figure~\ref{fig:overview} summarizes the OSplit workflow, from estimating the prefill and decode capacities for candidate parallelism configurations to selecting the GPU allocation and organizing the resulting model instances within an L1 domain.

\section{The GPU Capacity Model}
\label{sec:model}

This section derives the two unknown capacities in Equation~(\ref{eq:problem}) $c_{\mathrm{PF}}$ and $c_{\mathrm{DC}}$, for one candidate choice of parallelism degrees; \S~\ref{sec:prov} evaluates every valid candidate and selects the optimal degrees. We present the analytical formulation using the NVIDIA H100 GPU as a reference architecture.

\subsection{Capacity}
\label{sec:model:cap}

In steady state, both pools serve the same stream of requests, so we measure both capacities in requests per second, the unit in which their throughputs must match. A request gives the two pools different amounts of work, $I$ prompt tokens in prefill and $\bar O$ generated tokens in decode, so their tokens per second differ and cannot be directly compared.
%\sout{We measure capacity in requests per second rather than tokens per second because every request passes through each pool  once, so the number of requests per second is directly comparable between the two pools} \ash{this is not true}.\sout{A request costs} \ash{costs: RP} \sout{the two pools different numbers of tokens, $I$ in prefill and $\bar O$ in decode, so the tokens per second of the two pools cannot be compared directly.}\ash{hence costs are not the same... }
A prefill instance takes a batch of $B_{\mathrm{PF}}$ prompts, processes them in one iteration of duration $\tau_{\mathrm{PF}}$, and
then starts the next batch, so it completes $B_{\mathrm{PF}}$ requests every $\tau_{\mathrm{PF}}$. A decode instance holds
$B_{\mathrm{DC}}$ requests, each of which stays for $\bar O$ iterations of duration $\tau_{\mathrm{DC}}$, so the instance completes $B_{\mathrm{DC}}$ requests every $\bar O\tau_{\mathrm{DC}}$ (Equations~(\ref{eq:tpf}) and~(\ref{eq:tdc})). Both statements follow from Little's
law~\cite{little1961}: the completion rate equals the number of requests held, divided by the time each one spends. Hence per GPU,
\begin{equation}
  c_{\mathrm{PF}} = \frac{B_{\mathrm{PF}}}{G_{\mathrm{PF}}\,\tau_{\mathrm{PF}}},
  \qquad
  c_{\mathrm{DC}} = \frac{B_{\mathrm{DC}}}{G_{\mathrm{DC}}\,\bar O\,\tau_{\mathrm{DC}}}.
  \label{eq:cap}
\end{equation}
To evaluate Equation~(\ref{eq:cap}), we need the batch size (\S~\ref{sec:model:batch}) and the iteration time (\S~\ref{sec:model:phase}) of each pool.

\subsection{Batch sizes}
\label{sec:model:batch}

A prefill batch is limited by the number of tokens processed in one iteration, and a decode batch by memory capacity, since a decode instance keeps the KV cache of the entire prompt in GPU memory until its last token is generated.

In vLLM, the prefill token budget $T_b$ limits the number of tokens processed in a single iteration. Each model copy batches complete prompts up to this token limit, subject to the available GPU memory for their KV cache. Since each prompt contains $I$ tokens and an instance has $g_d^{\mathrm{PF}}$ copies of the model, the prefill batch size is: 
\begin{equation}
    B_{\mathrm{PF}} = g_d^{\mathrm{PF}}\min\Big(\Big\lfloor \frac{T_b}{I} \Big\rfloor,\ \Big\lfloor \frac{U M_{\mathrm{HBM}} - W - M_{\mathrm{res}}}{\kappa\, I} \Big\rfloor\Big) .
  \label{eq:bpf}
\end{equation}
The second term is the number of prompts whose KV caches fit in the memory that is left once the model weights and the framework's reserved memory are allocated; $W$, $M_{\mathrm{res}}$ and $\kappa$ are defined below. We set $T_b\ge I$, ensuring that each prompt fits within the token budget and each model copy can process at least one prompt per iteration.

The decode batch size depends on how much GPU memory is available for the KV cache. The parameter \texttt{gpu\_memory\_utilization} sets the fraction $U$ of GPU memory $M_{\mathrm{HBM}}$ that can be used. The model weights occupy $W$ bytes on each GPU, and the framework reserves $M_{\mathrm{res}}$ for its own buffers (peak activation memory, CUDA graph memory, context and communication buffers). Since each request requires $\kappa S$ bytes on average, the remaining memory determines how many requests each model copy can hold. For an instance with $g_d^{\mathrm{DC}}$ copies, the largest batch whose KV cache fits is
\begin{equation}
  B_{\mathrm{DC}} = g_d^{\mathrm{DC}}\Big\lfloor \frac{U M_{\mathrm{HBM}} - W - M_{\mathrm{res}}}{\kappa\, S} \Big\rfloor.
  \label{eq:bdc}
\end{equation}
%\sout{where the values of $W$ and $\kappa$ depend on the chosen decode parallelism degrees.} 
$W$ and $\kappa$ depend on the candidate parallelism. The parallelism is an output: the model evaluates Equation~(\ref{eq:bdc}) for every candidate, and \S~\ref{sec:prov} chooses the optimal. %\S~\ref{sec:prov:par} chooses the decode batch at or below $B_{\mathrm{DC}}$. %\ag{i thought this was an output of the system? needs discussion. } \ash{not yet resolved}

%Since the cache of a request grows by one token per decode iteration, its length increases from $I$ tokens 
%\ash{in this I, you do not consider batch size? to be it should be IB to (I+O)B???? and O should be O of I}

The output length $O(I)$ depends on the input prompt and model characteristics, making it difficult to predict before execution.
We therefore model $O(I)$ as a random variable with distribution $\hat O$ and mean $\bar O=\mathrm{E}[O(I)]$, estimated from the workload.
A request's KV cache grows from $I$ to $I+O(I)$ tokens during decode.
%Let $S$ denote its cache length.
At full load, decode batch slots remain occupied by successive requests.
A randomly observed slot is thus more likely to contain a longer response, with its position uniformly distributed over that response~\cite{gallager1996dsp}.
Consequently, the mean cache length $S$ is ($c_O$ denotes the coefficient of variation of $O(I)$):
\begin{subequations}
\label{eq:s}
\begin{align}
  S &= I + \frac{\mathrm{E}\big[O(I)^2\big]}{2\,\mathrm{E}\big[O(I)\big]} = I + \frac{\bar O}{2}\big(1+c_O^2\big), \label{eq:sa}\\
  S &\approx I + \frac{\bar O}{2}. \label{eq:sb}
\end{align}
\end{subequations}
For each KV head in every layer, the cache stores a key and a value vector of size $d_h$. The number of KV heads per GPU depends on the candidate TP degree $g_t$. 
In vLLM, $g_t$ must divide the number of query heads $H_q$, and either $g_t$ must divide $H_{kv}$, or $H_{kv}$ must divide $g_t$. The KV heads are distributed evenly among the $g_t$ GPUs or each GPU holds one KV head and each head is replicated on $g_t/H_{kv}$ GPUs. With $e_{kv}$ bytes per element (one for FP8), a token
occupies $\kappa$ bytes on one GPU:
\begin{equation}
  \kappa = 2\,L\,h_{kv}\,d_h\,e_{kv},
  \qquad
  h_{kv} = \max\left(1,\frac{H_{kv}}{g_t}\right)
  \label{eq:kappa}
\end{equation}
Equation~(\ref{eq:bdc}) shows that the optimal GPU split depends on the serving framework, model architecture, and hardware characteristics. A higher TP degree raises the number of requests each GPU holds, since each GPU stores a smaller share of the weights and of every request's
cache, but only until each GPU holds a single KV head.

\subsection{Prefill and decode times}
\label{sec:model:phase}

The prefill and decode iteration times $\tau_{\mathrm{PF}}$ and $\tau_{\mathrm{DC}}$ are the inputs to Equation~(\ref{eq:cap}). Together with the batch sizes, these two determine the capacity of each pool and consequently, how GPUs should be split between prefill and decode. In prefill, the model processes the entire batch at once, while in decode it generates one new token for each request in a batch (iteration). Both phases perform the same sequence of operations, layer by layer, but the amount of work in each operation differs. We first express the iteration time of each phase in terms of these operations, then derive the time taken by each operation in the next subsection.

A forward pass repeats the layer computation $L$ times and then runs the final normalization and LM head to obtain scores for the next token. With $b$ requests on the GPU:
\begin{align}
  \tau_{\mathrm{PF}} &= L\,\big(T_{\mathrm{layer}}(bI) + T_{\mathrm{kv}}\big) + T_{\mathrm{head}}, \label{eq:tpf}\\
  \tau_{\mathrm{DC}} &= L\,T_{\mathrm{layer}}(b) + T_{\mathrm{head}} + T_{\mathrm{samp}} + T_{\mathrm{host}}. \label{eq:tdc}
\end{align}
Here $T_{\mathrm{kv}}$ is the per-layer time to write the prompt's keys
and values to the cache; $T_{\mathrm{head}}$ covers the final
normalization, the LM head computation and the collective across the $g_t$ GPUs. $T_{\mathrm{samp}}$ is the time for sampling the next token, and $T_{\mathrm{host}}$ is vLLM's fixed CPU time per decode step, measured once per GPU. The LM head works on one row per request in both phases, since only the last prompt position produces a token.

The $g_d$ data parallel copies of the model process independent batches of $b$ requests in parallel. The copies run the same operations on batches of the same size simultaneously and exchange data only in the token exchange of an MoE layer,  so we model the time taken by one copy. Within each copy, the $g_t$ TP GPUs work on different shards of the model, while executing the same series of operations. In vLLM the kernels of a layer run one after another on each GPU, so their times add up (on GB300, in decode, DeepSeek's shared expert overlaps the routed experts when a GPU holds at most 256 tokens), and a layer on one GPU serving $b$ requests takes: 
\begin{equation}
  T_{\mathrm{layer}}(R) = 2\,T_{\mathrm{norm}}(R) + T_{\mathrm{ATT}}(R) + T_{\mathrm{FFN}}(R),
  \label{eq:layer}
\end{equation}
where,
\begin{align*}
  T_{\mathrm{ATT}} &= T_{\mathrm{GEMM}}^{QKV} + T_{\mathrm{attn}} + T_{\mathrm{GEMM}}^{O} + T_{\mathrm{AR}}(g_t),\\
  T_{\mathrm{FFN}}^{\mathrm{dense}} &= T_{\mathrm{GEMM}}^{\uparrow} + T_{\mathrm{GEMM}}^{\downarrow} + T_{\mathrm{AR}}(g_t),\\
    T_{\mathrm{FFN}}^{\mathrm{MoE}} &= T_{\mathrm{AG}}(G) + T_{\mathrm{RS}}(G) + \Lambda\,\big[T_{\mathrm{exp}}^{\uparrow} + T_{\mathrm{exp}}^{\downarrow}\big].
\end{align*}
Here, $R$ is the number of token rows processed by a layer. The layer time $T_{\mathrm{layer}}$ is the sum of the attention block, the feed-forward block, and two normalizations, one before each block. The attention time $T_{\mathrm{ATT}}$ accounts for the QKV projection, attention computation, output projection, and an all-reduce to combine results across the $g_t$ TP GPUs.
The FF time depends on the model architecture. In a dense model,
$T_{\mathrm{FFN}}^{\mathrm{dense}}$ includes the fused gate-and-up projection,
the down projection, and an all-reduce. In an MoE model,
$T_{\mathrm{FFN}}^{\mathrm{MoE}}$ accounts for dispatching tokens to their
selected experts, computing the expert outputs, and returning the results.
As in vLLM, the dispatch is an all-gather and the return a reduce-scatter over the $G$ GPUs. $\Lambda$ is the expected load of the busiest GPU relative to the mean (\S~\ref{sec:model:ops}).

The phases differ mainly in the amount of work %\sout{The main difference between the phases is the amount of work} \ag{amount or type or both???} 
each operation receives. During prefill, each prompt contributes $I$ token
rows, so a batch of $b$ requests gives $R=bI$ rows to the layer
operations. During decode, each request contributes just one new token,
giving $R=b$ rows. 
Prefill attention relates tokens within each prompt,
so its work grows quadratically with $I$. Decode attention instead
reads each request's KV cache, whose average length is $S$. Thus,
prefill time depends on the batch size and prompt length, while decode
time depends on the batch size and cache length.

The equations above express the prefill and decode iteration times in terms of individual
operation times. We next derive these times from the GPU's computational, memory and communication.

\subsection{Operation times}
\label{sec:model:ops}

We estimate the time of each operation from the resources used: the tensor cores, the L2 cache, HBM and the
links between GPUs. The model uses the GPU's specifications and some values measured once per GPU type, such as achieved HBM bandwidth and vLLM's fixed host cost per step, rather than profiling individual operations. For GEMMs and attention we build on the tile-level formulation of
TileSight~\cite{tilesight}, which we extend to FP8 arithmetic~\cite{fp8formats}, different tile sizes and split-K factors, and grouped GEMMs for MoE experts. This subsection outlines how each operation is modeled.

\noindent\textbf{GEMMs.} A GEMM multiplies an $M\times K$ matrix of activations by a $K\times N$ weight matrix. The GPU computes this in tiles of $t_m\times t_n$ elements. Each block of threads computes one tile, processing the $K$ dimension in chunks of $t_k$. With split-K, the $K$ dimension is divided among $n_s$ blocks whose
partial results are combined at the end, as described in~\cite{streamk}. The kernel therefore launches $n_{\mathrm{cta}}$ blocks of $n_k$ steps each,
\begin{equation}
  n_{\mathrm{cta}} = n_s\Big\lceil\frac{M}{t_m}\Big\rceil\Big\lceil\frac{N}{t_n}\Big\rceil,
  \qquad
  n_k = \Big\lceil\frac{K}{n_s\,t_k}\Big\rceil .
  \label{eq:tiles}
\end{equation}
The GPU schedules at most one block on each of the $N_{\mathrm{SM}}$ SMs at a time, so $p=\min(n_{\mathrm{cta}},N_{\mathrm{SM}})$ blocks run at once, the blocks run in $n_w=\lceil n_{\mathrm{cta}}/N_{\mathrm{SM}}\rceil$ waves, and the last wave may be partly empty. In CUDA terms, the kernel is launched with a grid of $n_{\mathrm{cta}}$ thread blocks; the index (\texttt{blockIdx.x}, \texttt{blockIdx.y}) of a block selects the output tile it works on, and \texttt{blockIdx.z} selects its split-K slice. In every step, each block loads $v_a=t_mt_ke_w$ bytes of activations and $v_b=t_kt_ne_w$ bytes
of weights, where $e_w$ is the number of bytes per element. Some of this data may be present in L2, while the rest must be fetched from HBM. The bytes fetched from HBM are $v_{\mathrm{miss}}=v_a(1-h_A)+v_b(1-h_B)$,
where $h_A$ and $h_B$ are the fractions of activation and weight data found in L2. With $p$ blocks running concurrently, the per step time is determined by whichever of the three resources (tensor cores, L2 cache or HBM) is the bottleneck:
\begin{equation}
  \tau_{\mathrm{step}} = \max\!\Big(\frac{2\,t_m t_n t_k}{F/N_{\mathrm{SM}}},\; \frac{p\,(v_a+v_b-v_{\mathrm{miss}})}{\beta_{L2}},\;\frac{p\,v_{\mathrm{miss}}}{\beta_H}\Big),
  \label{eq:lanes}
\end{equation}
where $F/N_{\mathrm{SM}}$ is the compute rate of one SM. 

A wave also incurs overhead when the kernel pipeline fills and the output is written. During the first $\delta$ steps, where $\delta$ is the kernel pipeline depth, the pipeline is still loading data and is limited by memory bandwidth. The remaining steps run at $\tau_{\mathrm{step}}$, and the wave ends
by writing the output tiles, which takes $\tau_e$. The time of one wave, $T_{\mathrm{wave}}$, is then:
\begin{equation}
  T_{\mathrm{wave}} = \delta\,\tau_{\mathrm{mem}} + \max(n_k-\delta,0)\,\tau_{\mathrm{step}} + \tau_e,
  \label{eq:wave}
\end{equation}
with $\tau_{\mathrm{mem}}$ the larger of the two memory terms of Equation~(\ref{eq:lanes}). The total GEMM time is obtained by adding the times of the $n_w-1$ full waves and a final wave along with a reduction pass when \emph{split-K} is used. The model evaluates the tile sizes that fit in shared memory and the available \emph{split-K} factors, and takes the configuration with the lowest estimated execution time as $T_{\mathrm{GEMM}}$.

We estimate $h_A$ and $h_B$ from the order in which the kernel assigns output tiles to blocks. A tile misses on its first access, and a subsequent access hits only if the tile remains in L2, which depends on how many other tiles were loaded. We compute this probability using the stack-distance model of TileSight~\cite{tilesight}, which models L2 as a set-associative LRU cache with tiles mapped to cache sets at random.

The model captures the difference between prefill and decode from the amount of work each GEMM receives, rather than assuming one is compute-bound and the other memory-bound. Decode processes few rows, so reading the model weights accounts for most of the bytes a GEMM operation moves, and HBM bandwidth becomes the potential bottleneck for this operation. Prefill processes many rows, making tensor-core throughput the potential bottleneck. A conventional roofline~\cite{roofline} estimate can identify these broad trends, but it does not capture how much data is served from L2 cache or the extra time caused by a partially filled final wave. 

\noindent\textbf{Expert GEMMs.} An MoE layer replaces one large FF GEMM with many small ones: each GPU holds $E/G$ experts and runs a separate matrix product for each expert. We estimate the time of these products using the same per-step model as a dense GEMM (Equation~(\ref{eq:lanes})), with the tile sizes from vLLM's expert kernels.

In prefill, a batch contains $B_{\mathrm{PF}}I$ token activations each routed to $k$ of the $E$ experts. So each expert receives about $B_{\mathrm{PF}}Ik/E$ tokens on
average. The actual distribution varies across batches because routing is stochastic and some experts are selected more often than others. Since the GPUs must finish their work before the layer can proceed, the GPU receiving the most tokens determines the execution time. We account for this imbalance using $\Lambda$,
the expected load of the busiest GPU relative to the mean, whose value depends on the number of tokens and the unevenness of the distribution.

During decode, each request contributes only one token/iteration, so a batch uses only a subset of the experts. For a batch of $B_{\mathrm{DC}}$
requests, the expected number of active experts per GPU is:
\begin{equation}
  a = \frac{E}{G}\Big(1-\Big(1-\frac{k}{E}\Big)^{B_{\mathrm{DC}}}\Big).
  \label{eq:active}
\end{equation}
The decode iteration therefore needs to read the weights only for the experts activated by the batch. Since these weight reads account for much of the data moved during decode, the number of active experts plays a key role in determining the layer's execution time.

\textit{Dynamics}: Prefill and decode pull the sizes in opposite directions. In prefill, spreading the experts across more GPUs leaves fewer experts on each GPU and can increase load imbalance, raising $\Lambda$. Smaller instances can therefore be more efficient for prefill. But for decode, larger instances reduce the number of active expert weights each GPU must read per step and can leave more memory available for the KV cache. The model can thus favor smaller instances for prefill and larger ones for decode.
%\sout{The two phases pull the instance size in opposite directions} \ag{elaborate on this}. \sout{In prefill, a larger $G$ makes the busiest GPU further from the mean and so raises $\Lambda$, while in decode, it lowers the number of expert weights each GPU must read per step.}

\noindent\textbf{Attention.} During decode, each request reads its KV cache of \(S\) tokens. When the batch is small, this work may not be enough to keep all the GPU's SMs busy. The kernel can split each cache into \(\sigma\) parts and process them in parallel across different SMs, but the partial results must be combined. Without splitting, the kernel runs one block per request and KV head, $b\,h_{kv}$ blocks in total, which can be fewer than the $N_{\mathrm{SM}}$ SMs. Splitting each cache into $\sigma$ parts gives $\sigma b\,h_{kv}$ blocks. The model takes $\sigma$ from the kernel's own split rule (FA-3 \cite{flashattention3} on H100) and runs the blocks in waves over the SMs: 
\begin{equation}
    T_{\mathrm{attn}}^{\mathrm{DC}} = \Big\lceil\frac{\sigma\,b\,h_{kv}}{N_{\mathrm{SM}}}\Big\rceil
  \max\Big(t_{\mathrm{comp}},\;\frac{v_{\mathrm{blk}}}{\beta_H},\;t_{\mathrm{lat}}\Big) + t_{\mathrm{merge}}(\sigma),
  \label{eq:attndc}
\end{equation}
where a block reads $v_{\mathrm{blk}}$ bytes of keys and values at the achieved bandwidth $\beta_H$, $t_{\mathrm{comp}}$: compute time on the tensor cores, special functions, and the CUDA
cores, $t_{\mathrm{lat}}$: latency of its dependent loads, and $t_{\mathrm{merge}}$:  cost of combining the partial results. On GB300, the MLA decode kernels are modeled similarly, each with its own split rule and merge.

During prefill, every token attends to itself and the preceding tokens in the prompt. The kernel processes queries in blocks of $t_m$ rows, and for each query block, it processes the blocks of $t_n$ keys that are before it, one key-block update at a time. On average, a query block makes $J_{\mathrm{eff}}$ such updates. Each update loads a block of keys and values, multiplies queries by keys on the tensor cores, computes the softmax on the special-function cores and CUDA cores, and multiplies the resulting attention weights by the values. Since the products and the softmax run on different units, consecutive updates can overlap, so their time depends on how the actions are ordered. 
Figure~\ref{fig:attn} illustrates the execution of consecutive key-block updates and the overlap of memory transfers, tensor-core operations, and softmax computations.
We find the shortest order of these actions that their dependencies allow, both for one update ($T_1$) and for two consecutive updates ($T_2$). The total attention time is then: 
\begin{equation}
  T_{\mathrm{attn}}^{\mathrm{PF}} = \min_{t_m,t_n}\Big[T_1 + (J_{\mathrm{eff}}-1)\,\max\big(T_2-T_1,\,u_{\max}\big) + T_{qo}\Big].
  \label{eq:attnpf}
\end{equation}
The first update costs $T_1$. Once consecutive updates overlap, each additional update adds $T_2-T_1$, subject to a lower bound of $u_{\max}$, the time of the busiest resource per update. $T_{qo}$ accounts for loading the queries and writing the output.

\input{fig_attn_prefill2}

\noindent\textbf{Collectives.} Collective operations exchange data among GPUs in a series of stages. Each stage takes as long as the slowest transfer, and
each collective adds a fixed latency $\alpha$. We model total time as:
\begin{equation}
  T_{\mathrm{coll}} = \alpha + \sum_{\text{stages}}\ \max_{\text{senders, links}} \frac{\text{bytes}}{\beta_{\text{NV}}} .
  \label{eq:coll}
\end{equation}
For an all-reduce of $X$ bytes over $g$ GPUs within an L1 domain, the ring algorithm takes $2(g-1)$ stages, each sends $X/g$ bytes per stage. So
$T_{\mathrm{AR}}=\alpha+2(g-1)X/(g\,\beta_{NV})$.  An all-gather or a reduce-scatter has half the stages of an all-reduce, while an all-to-all is a single stage in
which each GPU sends $(g-1)X/g$ bytes. These communication costs arise from TP and EP. 

\noindent\textbf{Other operations.} For normalization, KV-cache writes, sampling and other element-wise operations, we estimate execution time from the amount of data transferred and the GPU’s HBM BW.

This completes the model. The operation times of this subsection add up to the iteration times $\tau_{\mathrm{PF}}$ and
$\tau_{\mathrm{DC}}$ (\S~\ref{sec:model:phase}); together with the batch sizes of \S~\ref{sec:model:batch},
they give the two capacities of Equation~(\ref{eq:cap}), which are the only unknowns in the provisioning problem of
Equation~(\ref{eq:problem}). 

\section{The OSplit Method}
\label{sec:prov}

This section solves the provisioning problem in Equation~(\ref{eq:problem}). From the prefill and decode capacities computed in \S~\ref{sec:model}, for each candidate parallelism, we derive the ideal GPU split and throughput per GPU, choose the best parallelism for each pool, and determine the number of instances to deploy in an L1 domain of $N$ GPUs.

\subsection{The balanced split}
\label{sec:prov:balance}

For a fixed choice of parallelism, throughput is limited by whichever pool has the lower capacity. Allocating more GPUs to that pool increases throughput until the two pool capacities match and our goal is to compute the optimal match from a throughput perspective. When the number of GPUs in the L1 domain is such that the optimal allocation leads to an integer prefill and decode (when ceiled up), then we term this allocation as \textit{feasible}. For example if the prefill to decode ratio is 1.5, and the L1 domain has 72 GPUs then the allocation is infeasible (43.2 prefill and 28.8 decode GPUs), else if it is 1.4 then the allocation is feasible (42 and 30 GPUs). A feasible case satisfies $m+n=N$, with minimal slack (up to 1/2 GPU in either group lying unused in SM+HBM), and $m\,c_{\mathrm{PF}}=n\,c_{\mathrm{DC}}$. We can express this balance in terms of the GPU cost of serving a request in each phase. Let: 
\begin{equation}
  P = \frac{G_{\mathrm{PF}}\,\tau_{\mathrm{PF}}}{B_{\mathrm{PF}}},
  \qquad
  D = \frac{G_{\mathrm{DC}}\,\tau_{\mathrm{DC}}}{B_{\mathrm{DC}}}
  \label{eq:PD}
\end{equation}
be the GPU-seconds a request needs in prefill and the GPU-seconds an output token needs in decode, where \S~\ref{sec:prov:alg} chooses the decode batch $B_{\mathrm{DC}}$ to minimize $D$. By
Equation~(\ref{eq:cap}), $c_{\mathrm{PF}}=1/P$ and $c_{\mathrm{DC}}=1/(\bar O D)$, so the ratio of prefill to decode GPUs that maximizes throughput, and the resulting throughput per GPU are: 
\begin{equation}
  r^* = \frac{m}{n} = \frac{P}{\bar O\,D},
  \qquad
  \frac{\phi^*}{N} = \frac{1}{P+\bar O\,D}.
  \label{eq:rstar}
\end{equation}
The total GPU cost of serving a request is $P+\bar O D$, giving an ideal throughput per GPU of $1/(P+\bar O D)$.

\subsection{Choosing the parallelism}
\label{sec:prov:par}

By Equation~(\ref{eq:rstar}), the throughput per GPU depends on the parallelism only through the sum $P+\bar O D$.
$P$ depends only on the prefill parallelism $(g_t^{\mathrm{PF}},g_d^{\mathrm{PF}})$ and $D$ only on the decode
parallelism $(g_t^{\mathrm{DC}},g_d^{\mathrm{DC}})$. The best prefill parallelism
minimizes $P$, and the best decode parallelism minimizes $D$. $r^*$ follows from the two minima. Since $P$ and $D$ depend on separate parallelism choices, we can find the continuous optimum by minimizing them independently. Integer instances can change which pair is best, so \S~\ref{sec:prov:alg} searches over all pairs for which one instance of each pool fits in the $N$ GPUs.

The remaining constraints of Equation~(\ref{eq:problem}) decide which parallelisms are candidates. A pair
$(g_t,g_d)$ is a candidate iff: 

\begin{enumerate}[nosep]
  \item  $H_q$ is perfectly divisible by $g_t$, and similarly, $H_{kv}$ is perfectly divisible by $g_t$  or $H_{kv}$ divides $g_t$. 
  
  \item the instance fits in memory: for prefill, each copy of the model holds at least one prompt, $B_{\mathrm{PF}}\ge g_d$ in Equation~(\ref{eq:bpf}); for decode, each copy of the model holds at least one request, $B_{\mathrm{DC}}\ge g_d$ in Equation~(\ref{eq:bdc}).
  \item for decode, if a limit $\tau$ on the time between tokens is set, the limit is met at some batch of at least 1 request per copy. 
\end{enumerate}

The time-between-tokens constraint applies only to decode. As the decode batch grows, each iteration takes longer because the GEMMs process more rows and attention must read more KV-cache. The constraint therefore limits the batch to:
\begin{equation}
  B_\tau = \max\bigl\{B\le B_{\mathrm{DC}} : g_d \mid B,\ \tau_{\mathrm{DC}}(B/g_d)\le\tau\bigr\}.
  \label{eq:btbt}
\end{equation}
The decode pool then uses the batch at or below $B_\tau$ that gives the lowest $D$.

\begin{algorithm}[t]
  \caption{OSplit in an L1 domain of $N$ GPUs.}
  \label{alg:prov}
  \begin{algorithmic}[1]
    \REQUIRE Model, GPU, framework and workload parameters; L1 domain size $N$; optional TBT limit $\tau$
    \ENSURE Continuous split $r^*$ and throughput $\phi^*/N$; for $N$ GPUs, the best pair of parallelisms, the pool
      sizes and throughput $\phi_{\mathrm{int}}$
    \STATE $\mathcal{C}_{\mathrm{PF}},\mathcal{C}_{\mathrm{DC}}\leftarrow$ candidate parallelisms (\S~\ref{sec:prov:par})
    \FOR{each $s\in\mathcal{C}_{\mathrm{PF}}$}
      \STATE Compute $B_{\mathrm{PF},s}$; discard $s$ iff $B_{\mathrm{PF},s}<g_{d,s}$
      \STATE Compute $\tau_{\mathrm{PF},s}$
      \STATE $P_s\leftarrow G_s\tau_{\mathrm{PF},s}/B_{\mathrm{PF},s}$
    \ENDFOR
    \FOR{each $t\in\mathcal{C}_{\mathrm{DC}}$}
      \STATE $\bar B_t\leftarrow B_{\mathrm{DC},t}$ \COMMENT{memory ceiling}
      \IF{$\tau$ is specified}
        \STATE $\bar B_t\leftarrow\max\{B\le\bar B_t:\tau_{\mathrm{DC},t}(B)\le\tau\}$
      \ENDIF
      \STATE Discard $t$ if $\bar B_t<g_{d,t}$
      \FOR{each $B\le\bar B_t$ with $g_{d,t}\mid B$}
        \STATE $\delta_t(B)\leftarrow G_t\,\tau_{\mathrm{DC},t}/B$
      \ENDFOR
      \STATE $D_t\leftarrow\min_B\delta_t(B)$
    \ENDFOR
    \STATE $s^*\leftarrow\arg\min_{s}P_s$;\ \ $t^*\leftarrow\arg\min_{t}D_t$
    \STATE $r^*\leftarrow P_{s^*}/(\bar O D_{t^*})$;\ \ $\phi^*/N\leftarrow 1/(P_{s^*}+\bar O D_{t^*})$
    \STATE $\phi_{\mathrm{int}}\leftarrow 0$
    \FOR{each pair $(s,t)\in\mathcal{C}_{\mathrm{PF}}\times\mathcal{C}_{\mathrm{DC}}$}
      \STATE $c_{\mathrm{PF}}\leftarrow 1/P_s$;\ \ $c_{\mathrm{DC}}\leftarrow 1/(\bar O D_t)$
      \STATE $m^*\leftarrow N c_{\mathrm{DC}}/(c_{\mathrm{PF}}+c_{\mathrm{DC}})$;\ \ $m_c\leftarrow G_s\lfloor m^*/G_s\rfloor$
      \WHILE{$m_c c_{\mathrm{PF}}>n(m_c)\,c_{\mathrm{DC}}$}
        \STATE $m_c\leftarrow m_c-G_s$
      \ENDWHILE
      \FOR{each $m\in\{m_c,\,m_c+G_s\}$ with $n(m)\ge G_t$}
        \STATE $\phi\leftarrow\min\bigl(m c_{\mathrm{PF}},\,n(m)\,c_{\mathrm{DC}}\bigr)$
        \IF{$\phi>\phi_{\mathrm{int}}$}
          \STATE $\phi_{\mathrm{int}}\leftarrow\phi$; record $(s,t,m,n(m))$
        \ENDIF
      \ENDFOR
    \ENDFOR
    \STATE \textbf{return} $r^*$, $\phi^*/N$, the recorded $(s,t,m,n)$ and $\phi_{\mathrm{int}}$
  \end{algorithmic}
  \vspace{-3.75pt}
\end{algorithm}

\subsection{The provisioning procedure}
\label{sec:prov:alg}
Algorithm~\ref{alg:prov} brings together the capacity model and the
optimization results above to determine how an L1 domain should be
provisioned. It considers candidate parallelism configurations for
each pool, estimates their costs, and selects the configuration and
GPU allocation that maximize throughput.

A candidate is a pair $(g_t,g_d)$, specifying the tensor-and-data-parallel degrees of an instance. We consider only configs that satisfy the constraints in \S~\ref{sec:prov:par}. For dense models, we fix $g_d=1$, since DP copies can instead be treated as separate instances.
For MoE models, we also search over $g_d$, since it affects how
experts are distributed; we require $G$ to divide the
number of experts. The conditions differ between
prefill and decode, so we maintain separate candidate sets,
$\mathcal{C}_{\mathrm{PF}}$ and $\mathcal{C}_{\mathrm{DC}}$.
We use $s$ and $t$ to denote a prefill and a decode candidate,
respectively.

The procedure has three stages. First, it computes the cost of every candidate.
For a prefill candidate $s$, it computes the batch size $B_{\mathrm{PF},s}$ and the iteration time $\tau_{\mathrm{PF},s}$; their ratio, scaled by the instance size, is the GPU-seconds per request,
$P_s=G_s\tau_{\mathrm{PF},s}/B_{\mathrm{PF},s}$.
A prefill candidate is discarded if its weights and the activations of a full token budget do not fit in memory, or if a copy of the model cannot hold one prompt.
For a decode candidate $t$, memory sets the largest batch, and the candidate is discarded if that batch is less than one request per model copy. 
%\sout{If a limit on the time between tokens is set, the batch is further restricted to feasible values.} 
From the remaining batches, the procedure uses the one with the lowest GPU-seconds per output token, $D_t$.
Second, the candidates that minimize $P_s$ and $D_t$ independently
give the continuous optimum in Equation~(\ref{eq:rstar}). 
Finally, the procedure finds the best allocation for an L1 domain of
$N$ GPUs. Since each pool must contain an integer number of instances, assigning $m$ GPUs to prefill leaves the decode pool with the remaining GPUs that form complete instances, $n(m)=G_{\mathrm{DC}}\lfloor (N-m)/G_{\mathrm{DC}}\rfloor$. As $m$ increases, the prefill capacity rises while the decode capacity falls. Since the throughput is the smaller of these capacities, its maximum for a given pair of candidates occurs either at $m_c$ or at $m_c+G_{\mathrm{PF}}$, where $m_c$ is the largest multiple of  $G_{\mathrm{PF}}$ for which prefill capacity does not exceed decode capacity. For each such pair, the procedure starts at the continuous balance point $m^*=N c_{\mathrm{DC}}/(c_{\mathrm{PF}}+c_{\mathrm{DC}})$, steps down in increments of $G_{\mathrm{PF}}$ until it reaches $m_c$, evaluates the two allocations, and returns the configuration with the highest throughput, along with its parallelism choices and instance counts.

The capacity model is evaluated for each candidate and, on the decode side, for each allowed batch, and the integer search reuses these results. The procedure therefore derives the deployment configuration from the analytical model and GPU-specific parameters, without profiling each deployment.

\subsection{Sensitivity to capacity errors}
\label{sec:prov:err}

We quantify the throughput loss from choosing a GPU split that differs from the ideal ratio $r^*$. With continuous pool sizes, a split at ratio $r$ achieves the following fraction of optimal throughput:
\begin{equation}
  \frac{\phi(r)}{\phi^*} =
  \begin{cases}
    \dfrac{r}{r^*}\cdot\dfrac{1+r^*}{1+r}, & r\le r^*\ \ \text{(prefill limits)},\\[8pt]
    \dfrac{1+r^*}{1+r}, & r\ge r^*\ \ \text{(decode limits)}
  \end{cases}
  \label{eq:loss}
\end{equation}
The throughput loss depends on the direction of the error. If the ratio is too low by a relative amount $a$, throughput falls by approximately $a/(1+r^*)$, whereas a ratio that is high by $a$ reduces it by approximately $r^*a/(1+r^*)$. These losses are asymmetric: when $r^*<1$, underallocating GPUs to prefill is more costly than overallocating them.

This determines which errors of the capacity model matter. If the model predicts $P$ and $D$ with relative errors
$\epsilon_P$ and $\epsilon_D$, Equation~(\ref{eq:rstar}) gives the predicted ratio:
\begin{equation}
  \hat r = r^*\,\frac{1+\epsilon_P}{1+\epsilon_D}.
  \label{eq:rerr}
\end{equation}
An error common to both phases cancels, and it can change the predicted throughput but not the recommended split. Only an error that differs between prefill and decode moves the split, potentially reducing throughput according to Equation~(\ref{eq:loss}).

\section{Implementation and deployment with vLLM}
\label{sec:impl}
 
\subsection{The model}
\label{sec:impl:model}
 \input{tab_const}
 \input{fig_parity_all}
 
The analytical models are implemented in Python and executed on a CPU-based system equipped with an AMD Ryzen 7 7840HS processor and 16 GB of RAM. For each candidate configuration, it predicts the batch sizes and iteration times, breaking each iteration down into individual operations and identifying the resource that limits each one. This allows us to trace the prediction to the operations and hardware bottlenecks that determine performance. The model uses constants measured once for the GPUs, listed in Table~\ref{tab:const}.
 
\subsection{Deployment with vLLM}
\label{sec:impl:vllm}
 
Algorithm~\ref{alg:prov} produces a configuration for vLLM (0.26.0 on H100, 0.32.0 on GB300). For dense models, the TP degree $g_t$ is set through \texttt{-{}-tensor-parallel-size}. For MoE models, we also set \texttt{-{}-data-parallel-size} to $g_d$ and enable \texttt{-{}-enable-expert-parallel}, giving $g_d$ attention replicas of $g_t$ GPUs each and an EP group spanning all $G$ GPUs. The token budget $T_b$, memory fraction $U$, and decode batch size per model copy map to \texttt{-{}-max-num-batched-tokens}, \texttt{-{}-gpu-memory-utilization}, and \texttt{-{}-max-num-seqs}, respectively. 

Prefill and decode instances communicate through vLLM's NIXL connector~\cite{nixl}, which transfers each request's KV cache over NVLink. %A proxy routes each request to a prefill instance and then to a decode instance. 
Through NVLink bandwidths KV cache transfer time is $\sim$1\% of the compute time. Also, the transfer is overlapped by the next prefill and hence not exposed (\S~\ref{sec:eval}).

\subsection{Extension to GB300}
\label{sec:impl:gb}

We extend Algorithm~\ref{alg:prov} and the underlying math to a GB300 NVL72 rack with vLLM~0.32.0. The capacity model and provisioning search remain unchanged, except that decode batches are padded to the sizes for which vLLM captures CUDA graphs. The main extension is to the operation time models in \S~\ref{sec:model:ops}, which account for Blackwell's tensor cores and memory hierarchy, as well as the kernels selected by vLLM, including FA-4~\cite{flashattention4}. These include operators for MLA~\cite{deepseekv2}, sparse~\cite{deepseekv32} and linear attention~\cite{kimilinear}, FP4 weights~\cite{nvfp4}, wide expert parallelism~\cite{deepseekv3} and the launch cost of kernels executed outside CUDA graphs. The GEMM and attention models also account for Blackwell-specific features like tensor memory and two-tier L2 cache. Like H100, the hardware constants are from datasheets and microbenchmarks.

%We identify the kernel used for each operation from the vLLM~0.32.0 source, and when the default kernel is closed source, we estimate its cost using the nearest open kernel alternative. 

\section{Evaluation}
\label{sec:eval}

We test OSplit on three aspects: the KV-cache capacity, the prefill and decode step times, and the throughput of every split of the node between prefill and decode. All predictions in this section come from our analytical model. On H100, we measured 11 KV capacities and 100 step times (22 prefill, 78 decode) across five models, with one to eight GPUs per instance. The ISL ranged from 1024 to 32,768, decode batch sizes of 1 to 1000, and six split curves. For GB300, we measured 12 KV capacities and 336 steps (60 prefill and 276 decode) across six models.
 
\subsection{Setup}
\label{sec:eval:setup}

We evaluate five models: three dense models, Llama~3.1 8B, 70B and 405B~\cite{llama3,hfllama8bfp8,hfllama70bfp8,hfllama405bfp8}, and two MoE models, Qwen3-30B-A3B and Qwen3-235B-A22B~\cite{hfqwen30bfp8,hfqwen235bfp8}, which activate 8 of 128 experts per token~\cite{qwen3}.\\
\textbf{Hardware and software.} All measurements run on one node of eight H100 SXM GPUs connected by NVLink, and one GB300 NVL72. H100 runs on vLLM~0.26.0 and GB300 runs on vLLM~0.32.0.\\
\textbf{Models.} Llama~3.1~405B and Qwen3-235B-A22B cannot be split on one node since one 405B instance needs all eight GPUs, and one 235B prefill instance and one decode instance together need 12 GPUs. The split is measured for Llama~3.1~8B and Qwen3-30B-A3B with 1 GPU per instance. We use 1 GPU per Llama~8B and Qwen3-30B, 2 for 70B and 8 for 405B. On GB300 we evaluate Llama~3.1~70B, Qwen3-30B-A3B and four DeepSeek models (V3, V3.2, and R1 and V3.2 with NVFP4 weights) on 1 to 32 GPUs per instance. \\
\textbf{Workloads.} The split is measured at three workloads, with mean ISL/OSL: W1 (1,024/2,048), W2 (8,192/700) and W3 (32,768/256). Prompts are random token sequences of exact length; output lengths are drawn uniformly
within $\pm$15\% of the mean and generation ignores end-of-sequence tokens. \\
\textbf{Measurements.} We record 4 metrics. \emph{Memory:} the KV-cache capacity reported by vLLM  at start-up. \emph{Prefill iteration time:} we pause the server, queue a full batch of prompts and resume execution, so that the batch runs as one iteration. \emph{Decode iteration time:} the time per iteration once the batch has reached steady state.
\emph{Split throughput}\emph{:} completed requests per second over a steady window. \\
\textbf{Baselines and metrics.} The \emph{roofline}~\cite{roofline} gives each operation the larger of its FLOPs at the peak tensor-core rate and its HBM traffic at the nominal bandwidth. \emph{AIConfigurator}~\cite{aiconfigurator} predicts the split from kernel and communication times from a database measured on each GPU. 
For each quantity, we measure the MAPE and the median signed error. For the split, we measure if the model chooses the split that measured best, and the median error of its predicted throughput over all splits.
\input{tab_acc}

\subsection{Memory}
\label{sec:eval:mem}
The model predicts vLLM's reported KV-cache capacity with a MAPE of 3.5\% over 11 H100 configurations (0.9--6.7\% per model), and 5.0\% over 12 GB300 configurations (Table~\ref{tab:acc}). It accounts for the precision of each module as specified in the checkpoint and for memory reserved by vLLM. These are important since Meta's FP8 Llama~405B checkpoint, for example, keeps several modules in BF16~\cite{hfllama405bfp8}.

\subsection{Iteration times}
\label{sec:eval:iter}
Figure~\ref{fig:parity} compares the predicted and measured prefill and decode times.\\
\textbf{Prefill.} On H100, the model predicts prefill time with a MAPE of 9.0\%, compared with 57\% for the roofline (Table~\ref{tab:acc}, Figure~\ref{fig:parity}a). The error ranges from 0.7--15.8\% across the five models. The roofline errors range from 43\% to 75\% because it assumes peak tensor-core throughput for attention and fully occupied waves of thread blocks. For Llama~8B, the relative error stays between $-6.3\%$ and $-8.0\%$ as the ISL increases from 1K to 32K. When this error is similar across candidates within a phase, it scales their estimated capacities without changing which candidate is best. Its effect on the optimal split depends on how it differs from the decode error (Equation~\ref{eq:rerr}). On GB300 (Figure~\ref{fig:parity}c) the prefill MAPE is 19.9\% (12--26\% per model) against the roofline's 58\%. \\
\textbf{Decode.} On H100 the model predicts decode time with a MAPE of 9.4\% (7.2--11.0\% per model). In contrast, the roofline has a MAPE of 40\%. The roofline misses decode by 35--47\%, since at small batches a decode step is dominated by fixed costs from small kernels, attention call costs and the host time of each step, which no bandwidth or FLOP count captures. On GB300 (Figure~\ref{fig:parity}d) the decode MAPE for our model is 16.8\%, compared with 56\% for the roofline. 

\input{tab_regret}
\input{fig_split_C}
\subsection{OSplit in action}
\label{sec:eval:split}
 In all six cases, OSplit chooses the optimal split (Figure~\ref{fig:split}, Table~\ref{tab:regret}). For W2, our median throughput prediction error ranges from 8.4--13.1\%, compared with 14.1--30.0\% for AIConfigurator. For W3, AIConfigurator is closer (19.5--20.9\%, against 25.0--27.0\% for OSplit). Two properties of OSplit make this choice robust: (1) the split depends only on the ratio $P/(\bar O D)$, so an error that affects prefill and decode alike cancels (Equation~(\ref{eq:rerr})); (2) on 8 $\times$
GPUs, the split can only change 1 GPU at a time, so the remaining error in the ratio changes the choice only if it is large enough to move the ratio past the point where two neighboring splits give the same throughput. 
For W1, decode limits every split, so both OSplit and AIConfigurator choose the smallest prefill pool.
Note that both OSplit and AIConfigurator identify the measured optimal GPU split in all six cases (Table~\ref{tab:regret}), indicating that both approaches provide sufficiently accurate capacity estimates for selecting the best allocation for these workloads.
However, their underlying methodologies differ: AIConfigurator uses an offline-profiled operator-performance database and enumerates candidate worker allocations, whereas OSplit predicts phase capacities analytically and derives the optimal integer allocation from the structure of the throughput objective. 
Thus, OSplit achieves the same allocation decisions without requiring a platform-specific operator-latency database.

\subsection{Cost of the method}
\label{sec:eval:cost}
 
A full provisioning search for one split takes 0.39--0.82~sec on one CPU core. Computing the whole sweep of \S~\ref{sec:results} (1,189 candidates) takes 112~sec on 8 CPU cores. The routing tables for MoE models are computed just once.

\section{What the model tells a user}
\label{sec:results}
 
We now apply our analytical model to the above 5$\times$ LLMs across ISLs from 1,024 to 65,536 and OSLs of 256 to
2,048 tokens, and to the six GB300 models with $N=72$. We examine how the optimal provisioning changes across these settings, explain the factors driving changes, and discuss their deployment implications.
 
\input{fig_ratio}
 
\noindent\textbf{The split moves with the workload.} For a fixed model and output length, the best ratio of prefill to decode GPUs grows 3.3--4.9$\times$ on H100 from 1,024 input tokens to the longest considered input length (Figure~\ref{fig:ratio}), and for a fixed input length it falls 8.0--13.2$\times$ from 256 to 2,048 output tokens. This is due to Equation~(\ref{eq:rstar}), $r^*=P/(\bar OD)$. The prefill cost of a request grows with the input linearly through the GEMMs and quadratically through attention, $P\approx p_1I+p_2I^2$. The decode cost of a token has a fixed part $d_0$, mostly reading the weights, and a part that grows with the cache attention reads, $D\approx d_0+d_1S$. This closed form serves only to explain the trend, not to predict: with $p_1,p_2,d_0,d_1$ fitted to the model's own $P$ and $D$, it reproduces $r^*$ within 8.2\% for every model and workload, so the trend follows from the two cost structures rather than from details of any one model. A single ratio chosen once for all workloads loses 41--43\% of the throughput at its worst workload, for every model. Figure~\ref{fig:ratio}b shows OSplit's prediction for GB300.

\input{fig_hbm}

\noindent\textbf{What decode reads.} The share of the decode step spent in attention grows with the input length for every model (Figure~\ref{fig:hbm}): from 66\% to 90\% for Llama~8B, from 33\% to 65\% for Llama~70B and from 7\% to 17\% for Llama~405B. Llama~405B reads far more weights per cached token, so even at the longest input, most of its decode step is spent reading weights. When weight reads dominate, increasing the batch amortizes weight reads across more requests, reducing $D$. As the KV cache becomes the main source of traffic, each request adds its own cache reads, limiting the benefit of larger batches; $D$ also increases with context length. 
 
\section{Related Work}
\label{sec:related}
\noindent\textbf{Disaggregated LLM serving.}
DistServe~\cite{distserve}, Splitwise~\cite{splitwise}, TetriInfer~\cite{tetriinfer}, and Mooncake~\cite{Mooncake} demonstrate the benefits of disaggregating prefill and decode.
Subsequent work explores resource allocation and adaptation across these phases. 
DOPD~\cite{dopd} dynamically adjusts the prefill--decode ratio, while SplitScaling~\cite{SplitScaling} employs reinforcement learning for workload-aware scaling.
Evolving Infra~\cite{kang2026evolving} adapts compute allocation and KV-cache management for agentic workloads.

\noindent\textbf{Performance modeling and configuration optimization.}
Vidur~\cite{vidur} combines operator profiling and simulation to predict LLM serving performance.
LLMCompass~\cite{zhang2024llmcompass} models inference performance across hardware configurations, while TileSight~\cite{tilesight} develops a tile-level
analytical model of GPU computation, memory, and communication, some of which we use here.
AIConfigurator~\cite{aiconfigurator} combines analytical operator models with a calibrated performance database to optimize serving configurations.
For disaggregated inference, SARA~\cite{sara} models prefill, KV-cache transfer, and decode queueing to optimize SLO-constrained goodput, while Yan et al.~\cite{yan2026power} analytically model
serving capacity and power consumption for resource provisioning.
OSplit builds on hardware-level performance modeling to estimate prefill and decode capacities, accounting for MoE expert execution and serving-framework constraints. It then exploits the structure of the two-stage throughput objective to jointly select parallelism configurations and derive an exact integer GPU allocation under the analytical model.

\section{Conclusion} 
\label{sec:conc}
We present OSplit, a fast and efficient technique for computing the disaggregation ratio of prefill to decode workers. OSplit combines an analytical capacity model with a search algorithm for disaggregated split computation, and is tested on H100 and GB300 GPUs across dense and MoE models. In every case measured, OSplit identifies the best split without profiling any deployment, showing that provisioning for disaggregated inference can be derived from first principles.

\FloatBarrier

\bibliography{refs}
\bibliographystyle{ACM-Reference-Format}

\end{document}

%% file: fig_styles.tex
\usepackage{pgfplots}
\pgfplotsset{compat=1.18}
\usepgfplotslibrary{groupplots,fillbetween}
\definecolor{c8b}{HTML}{0072B2}\definecolor{c70b}{HTML}{E69F00}\definecolor{c405b}{HTML}{009E73}
\definecolor{cq30}{HTML}{CC79A7}\definecolor{cq235}{HTML}{D55E00}\definecolor{cdv}{HTML}{56B4E9}
\definecolor{cr1}{HTML}{882255}\definecolor{cv32f}{HTML}{44AA99}
\pgfplotsset{
  osplit/.style={font=\scriptsize, tick label style={font=\scriptsize}, label style={font=\scriptsize},
    legend style={font=\scriptsize, draw=none, fill=none}, legend cell align=left,
    axis line style={black!70}, tick style={black!70}, grid style={black!12}, ymajorgrids, xmajorgrids,
    every axis plot/.append style={line width=0.7pt}, scaled y ticks=false},
  m8b/.style={color=c8b, mark=*, mark size=1.2pt},
  m70b/.style={color=c70b, mark=square*, mark size=1.1pt},
  m405b/.style={color=c405b, mark=diamond*, mark size=1.5pt},
  mq30/.style={color=cq30, mark=triangle*, mark size=1.5pt},
  mq235/.style={color=cq235, mark=pentagon*, mark size=1.4pt},
  mdv3/.style={color=black, mark=*, mark size=1.1pt},
  mdv32/.style={color=cdv, mark=square*, mark size=1.1pt},
  mdr1/.style={color=cr1, mark=triangle*, mark size=1.5pt},
  mdv32f/.style={color=cv32f, mark=diamond*, mark size=1.5pt},
}

%% file: fig_overview_1.tex
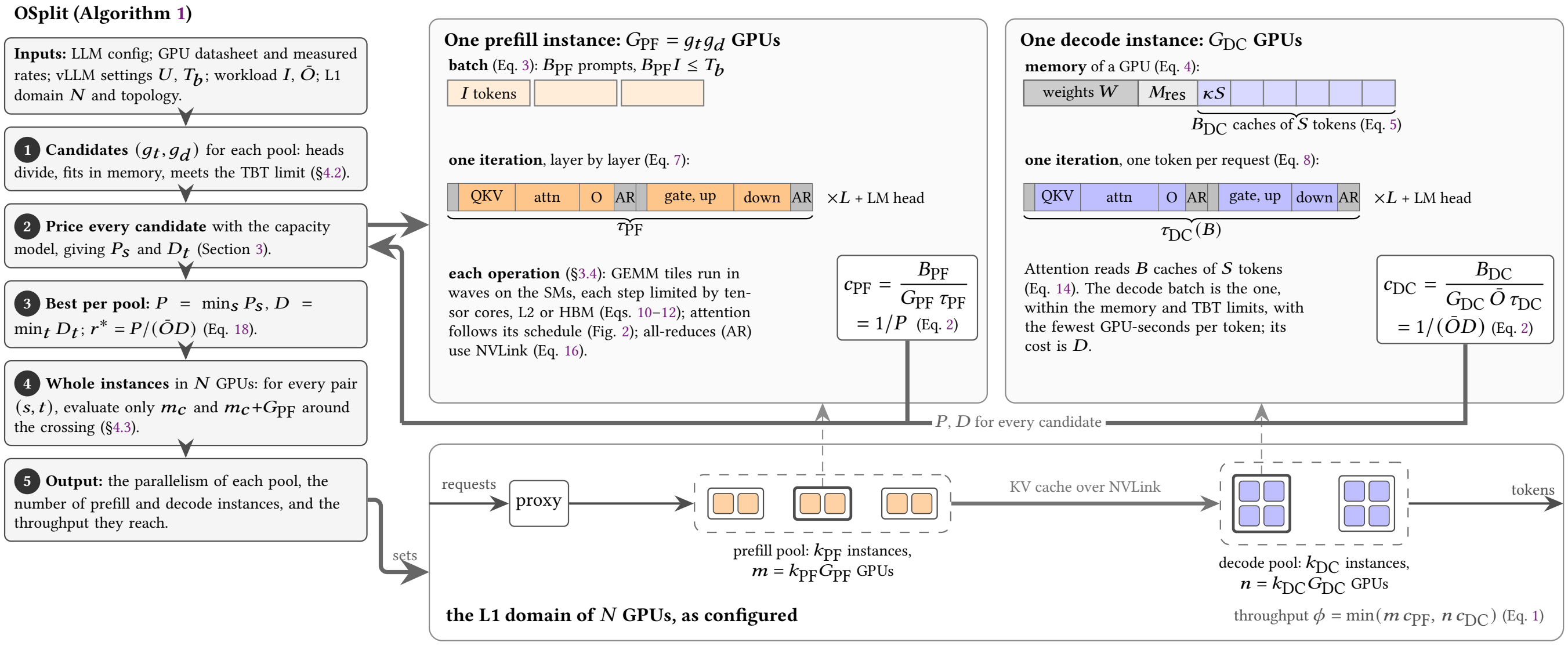
\begin{figure*}[t]
\centering
\resizebox{\textwidth}{!}{\begin{tikzpicture}[font=\scriptsize,>={Stealth[length=6pt, width=5pt]},
  alg/.style={draw=black!70,rounded corners=2pt,fill=black!4,align=left,inner sep=3pt,text width=3.75cm,font=\tiny},
  zoom/.style={draw=black!45,rounded corners=3pt,fill=black!2},
  box/.style={draw=black!65,rounded corners=1.5pt,fill=white,align=center,inner sep=2pt},
  stp/.style={circle,draw=black!80,fill=black!80,text=white,font=\tiny\bfseries,inner sep=1pt,minimum size=0.32cm},
  ref/.style={font=\tiny,black!60},
  arr/.style={->,black!70,line width=0.7pt},
  big/.style={-{Stealth[length=9pt,width=8pt]},black!60,line width=1.3pt},
  gpf/.style={draw=black!70,fill=orange!35,minimum size=0.22cm,inner sep=0pt,line width=0.3pt},
  gdc/.style={draw=black!70,fill=blue!25,minimum size=0.22cm,inner sep=0pt,line width=0.3pt},
  inst/.style={draw=black!70,rounded corners=1pt,inner sep=1.5pt},
  tc/.style={fill=orange!45,draw=black!60,line width=0.2pt},
  hb/.style={fill=blue!25,draw=black!60,line width=0.2pt},
  nv/.style={fill=black!25,draw=black!60,line width=0.2pt}]

\node[font=\scriptsize\bfseries,anchor=west] at (0,9.55) {OSplit (Algorithm~\ref{alg:prov})};
\node[alg,anchor=north west] (in) at (0,9.3) {\textbf{Inputs:} LLM config; GPU datasheet and measured rates; vLLM settings $U$, $T_b$; workload $I$, $\bar O$; L1 domain $N$ and topology.};
\node[alg,below=0.13cm of in] (s1) {\tikz[baseline=(n.base)]\node[stp,minimum size=0.27cm](n){1}; \textbf{Candidates} $(g_t,g_d)$ for each pool: heads divide, fits in memory, meets the TBT limit (\S\ref{sec:prov:par}).};
\node[alg,below=0.13cm of s1] (s2) {\tikz[baseline=(n.base)]\node[stp,minimum size=0.27cm](n){2}; \textbf{Price every candidate} with the capacity model, giving $P_s$ and $D_t$ (Section~\ref{sec:model}).};
\node[alg,below=0.13cm of s2] (s3) {\tikz[baseline=(n.base)]\node[stp,minimum size=0.27cm](n){3}; \textbf{Best per pool:} $P=\min_s P_s$, $D=\min_t D_t$; $r^*=P/(\bar O D)$ (Eq.~\ref{eq:rstar}).};
\node[alg,below=0.13cm of s3] (s4) {\tikz[baseline=(n.base)]\node[stp,minimum size=0.27cm](n){4}; \textbf{Whole instances} in $N$ GPUs: for every pair $(s,t)$, evaluate only $m_c$ and $m_c{+}G_{\mathrm{PF}}$ around the crossing (\S\ref{sec:prov:alg}).};
\node[alg,below=0.13cm of s4] (s5) {\tikz[baseline=(n.base)]\node[stp,minimum size=0.27cm](n){5}; \textbf{Output:} the parallelism of each pool, the number of prefill and decode instances, and the throughput they reach.};
\foreach \a/\b in {in/s1,s1/s2,s2/s3,s3/s4,s4/s5}{\draw[arr] (\a.south) -- (\b.north);}

\draw[zoom] (4.65,5.3) rectangle (10.75,9.5);
\node[anchor=north west,font=\scriptsize\bfseries] at (4.7,9.47) {One prefill instance: $G_{\mathrm{PF}}=g_tg_d$ GPUs};
\node[anchor=west,font=\tiny] at (4.75,8.97) {\textbf{batch} (Eq.~\ref{eq:bpf}): $B_{\mathrm{PF}}$ prompts, $B_{\mathrm{PF}}I\le T_b$};
\foreach \k in {0,1,2}{\draw[fill=orange!15,draw=black!50] (4.85+\k*0.95,8.55) rectangle ++(0.9,0.28);}
\node[font=\tiny] at (5.3,8.69) {$I$ tokens};
\node[anchor=west,font=\tiny] at (4.75,7.95) {\textbf{one iteration}, layer by layer (Eq.~\ref{eq:tpf}):};
\foreach \x/\w/\s/\t in {4.85/0.12/nv/,4.97/0.62/tc/QKV,5.59/0.70/tc/attn,6.29/0.38/tc/O,6.67/0.24/nv/AR,6.91/0.12/nv/,7.03/0.95/tc/{gate, up},7.98/0.62/tc/down,8.60/0.24/nv/AR}
  {\draw[\s] (\x,7.4) rectangle ++(\w,0.3); \node[font=\tiny] at (\x+\w/2,7.55) {\t};}
\node[font=\tiny,anchor=west] at (8.88,7.55) {$\times L$ + LM head};
\draw[decorate,decoration={brace,amplitude=2pt,mirror}] (4.85,7.35) -- node[below=-1pt,font=\tiny]{$\tau_{\mathrm{PF}}$} (8.84,7.35);
\node[anchor=north west,align=left,text width=3.3cm,font=\tiny] at (4.75,6.9) {\textbf{each operation} (\S\ref{sec:model:ops}): GEMM tiles run in waves on the SMs, each step limited by tensor cores, L2 or HBM (Eqs.~\ref{eq:tiles}--\ref{eq:wave}); attention follows its schedule (Fig.~\ref{fig:attn}); all-reduces (AR) use NVLink (Eq.~\ref{eq:coll}).};
\node[box,anchor=north east] (cpf) at (10.65,6.92) {$c_{\mathrm{PF}}=\dfrac{B_{\mathrm{PF}}}{G_{\mathrm{PF}}\,\tau_{\mathrm{PF}}}$\\[1pt]$=1/P$ \ {\tiny(Eq.~\ref{eq:cap})}};

\draw[zoom] (10.95,5.3) rectangle (17.05,9.5);
\node[anchor=north west,font=\scriptsize\bfseries] at (11.0,9.47) {One decode instance: $G_{\mathrm{DC}}$ GPUs};
\node[anchor=west,font=\tiny] at (11.05,8.97) {\textbf{memory} of a GPU (Eq.~\ref{eq:bdc}):};
\draw[fill=black!20,draw=black!60] (11.15,8.55) rectangle ++(1.25,0.28) node[midway,font=\tiny]{weights $W$};
\draw[fill=black!8,draw=black!60] (12.4,8.55) rectangle ++(0.65,0.28) node[midway,font=\tiny]{$M_{\mathrm{res}}$};
\foreach \k in {0,...,5}{\draw[fill=blue!15,draw=black!50] (13.05+\k*0.36,8.55) rectangle ++(0.36,0.28);}
\node[font=\tiny] at (13.23,8.69) {$\kappa S$};
\draw[decorate,decoration={brace,amplitude=2pt,mirror}] (13.05,8.5) -- node[below=-1pt,font=\tiny]{$B_{\mathrm{DC}}$ caches of $S$ tokens (Eq.~\ref{eq:s})} (15.21,8.5);
\node[anchor=west,font=\tiny] at (11.05,7.95) {\textbf{one iteration}, one token per request (Eq.~\ref{eq:tdc}):};
\foreach \x/\w/\s/\t in {11.15/0.12/nv/,11.27/0.5/hb/QKV,11.77/0.85/hb/attn,12.62/0.3/hb/O,12.92/0.24/nv/AR,13.16/0.12/nv/,13.28/0.8/hb/{gate, up},14.08/0.5/hb/down,14.58/0.24/nv/AR}
  {\draw[\s] (\x,7.4) rectangle ++(\w,0.3); \node[font=\tiny] at (\x+\w/2,7.55) {\t};}
\node[font=\tiny,anchor=west] at (14.86,7.55) {$\times L$ + LM head};
\draw[decorate,decoration={brace,amplitude=2pt,mirror}] (11.15,7.35) -- node[below=-1pt,font=\tiny]{$\tau_{\mathrm{DC}}(B)$} (14.82,7.35);
\node[anchor=north west,align=left,text width=3.15cm,font=\tiny] at (11.05,6.95) {Attention reads $B$ caches of $S$ tokens (Eq.~\ref{eq:attndc}). The decode batch is the one, within the memory and TBT limits, with the fewest GPU-seconds per token; its cost is $D$.};
\node[box,anchor=north east] (cdc) at (16.95,6.92) {$c_{\mathrm{DC}}=\dfrac{B_{\mathrm{DC}}}{G_{\mathrm{DC}}\,\bar O\,\tau_{\mathrm{DC}}}$\\[1pt]$=1/(\bar O D)$\ {\tiny(Eq.~\ref{eq:cap})}};
% =========================== the L1 domain (bottom right) ===========================
\draw[black!45,rounded corners=4pt,fill=white] (4.65,2.7) rectangle (17.05,4.85);
\node[anchor=south west,font=\scriptsize\bfseries] at (4.72,2.75) {the L1 domain of $N$ GPUs, as configured};
\node[box,minimum height=0.5cm] (px) at (5.85,4.2) {proxy};
\foreach \i in {0,1,2}{\node[inst] (pf\i) at (8.0+\i*0.95,4.2) {\tikz{\node[gpf] at (0,0){};\node[gpf] at (0.27,0){};}};}
\node[draw=black!55,dashed,rounded corners=3pt,inner sep=4pt,fit=(pf0)(pf2)] (pp) {};
\node[anchor=north,font=\tiny,align=center] at (pp.south) {prefill pool: $k_{\mathrm{PF}}$ instances,\\$m=k_{\mathrm{PF}}G_{\mathrm{PF}}$ GPUs};
\foreach \i in {0,1}{\node[inst] (dc\i) at (13.75+\i*1.15,4.2) {\tikz{\foreach \a in {0,1}{\foreach \b in {0,1}{\node[gdc] at (0.27*\a,0.27*\b){};}}}};}
\node[draw=black!55,dashed,rounded corners=3pt,inner sep=4pt,fit=(dc0)(dc1)] (dp) {};
\node[anchor=north,font=\tiny,align=center] at (dp.south) {decode pool: $k_{\mathrm{DC}}$ instances,\\$n=k_{\mathrm{DC}}G_{\mathrm{DC}}$ GPUs};
\draw[arr] (4.65,4.2) -- node[above,font=\tiny]{requests} (px.west);
\draw[arr] (px.east) -- (pp.west);
\draw[arr,very thick,black!55] (pp.east) -- node[above,font=\tiny]{KV cache over NVLink} (dp.west);
\draw[arr] (dp.east) -- (17.05,4.2) node[above left=-1pt,font=\tiny]{tokens};
\node[font=\tiny,anchor=south east,align=right,black!70] at (16.95,2.75) {throughput $\phi=\min(m\,c_{\mathrm{PF}},\,n\,c_{\mathrm{DC}})$ (Eq.~\ref{eq:problem})};
% zoom lines: the outlined instance of each pool is the one priced above
\draw[black!45,densely dashed,->] (pf1.north) -- (pf1.north |- 0,5.3);
\draw[black!45,densely dashed,->] (dc0.north) -- (dc0.north |- 0,5.3);
\draw[black!70,line width=0.8pt,rounded corners=1pt] (pf1.south west) rectangle (pf1.north east);
\draw[black!70,line width=0.8pt,rounded corners=1pt] (dc0.south west) rectangle (dc0.north east);
% =========================== connections ===========================
\draw[big] ([yshift=0.12cm]s2.east) -- ([yshift=0.12cm]s2.east -| 4.65,0);
\coordinate (jn) at (cdc.south |- 0,5.08); \draw[black!60,line width=1.3pt] (cpf.south) -- (cpf.south |- jn);
\draw[black!60,line width=1.3pt] (cdc.south) -- (jn) -- node[pos=0.42,fill=white,inner sep=1pt,font=\tiny]{$P$, $D$ for every candidate} (4.35,0 |- jn);
\draw[big,rounded corners=2pt] (4.35,0 |- jn) |- ([yshift=-0.12cm]s2.east);
\draw[big,rounded corners=2pt] (s5.east) -- (s5.east -| 4.12,0) |- node[above,font=\tiny,pos=0.75]{sets} (4.65,3.45);
\end{tikzpicture}}
\caption{OSplit on an L1 domain. Left: the OSplit Algorithm (Algorithm~\ref{alg:prov}, \S~\ref{sec:prov}). Top right: the GPU capacity model (\S~\ref{sec:model}), which gives each candidate's costs $P$ and $D$, shown here for the two outlined instances. Bottom: the resulting L1 domain, with 2-GPU prefill and 4-GPU decode instances.}
\label{fig:overview}
\end{figure*}

%% file: fig_attn_prefill2.tex
\begin{figure}[t]
  \centering
\begin{tikzpicture}[font=\fontsize{8}{9.6}\selectfont, >={Stealth[length=3.2pt]},
    act/.style={draw=black!65, line width=0.4pt, rounded corners=0.6pt},
    kk/.style={act, fill=blue!17},   vv/.style={act, fill=yellow!45},
    tc/.style={act, fill=orange!32}, sm/.style={act, fill=green!28},
    io/.style={act, fill=black!9},
    nxt/.style={postaction={pattern=north east lines, pattern color=black!55}},
    dag/.style={inner xsep=3pt, inner ysep=1.6pt, text height=8.6pt, text depth=3.4pt},
    dep/.style={->, line width=0.45pt},
    xdep/.style={->, densely dashed, line width=0.45pt, black!70},
    br/.style={decorate, decoration={brace, amplitude=2.5pt}, line width=0.4pt},
    guide/.style={densely dotted, black!55, line width=0.4pt},
    lbl/.style={font=\fontsize{8}{9.6}\selectfont\bfseries}]

\begin{scope}
  \node[lbl, anchor=west, inner xsep=0pt] at (0, 0.85) {(a)};
  \node[kk, dag, anchor=west] (lk) at (0, 0)  {load $K_j$};
  \node[tc, dag, right=0.28cm of lk] (qk)       {$Q_iK_j^{\!\top}$};
  \node[sm, dag, right=0.28cm of qk] (sm)       {softmax$_j$};
  \node[tc, dag, below=0.42cm of sm] (pv)       {$P_{ij}V_j$};
  \node[vv, dag] (lv) at (lk |- pv)             {load $V_j$};
  \draw[dep] (lk) -- (qk);
  \draw[dep] (qk) -- (sm);
  \draw[dep] (sm) -- (pv);
  \draw[dep] (lv) -- (pv);
  
  \draw[xdep] (sm.north) -- ++(0, 0.38) node[above, inner sep=1pt, font=\fontsize{7}{8.4}\selectfont]
        {softmax$_{j+1}$ (running max)};
  \draw[xdep] (pv.south) -- ++(0, -0.38) node[below, inner sep=1pt, font=\fontsize{7}{8.4}\selectfont]
        {$P_{i,j+1}V_{j+1}$ (adds to $O_i$)};
\end{scope}

\begin{scope}[shift={(5.55, 0.62)}]
  \def\c{0.27}
  \node[lbl, anchor=west] at (-0.62, 0.23) {(b)};
  % visited blocks (j < i): full; diagonal blocks: lower triangle only
  \foreach \r in {0,...,5} {
    \foreach \col in {0,...,5} {
      \ifnum\col>\r\else
        \fill[black!14] (\col*\c, -\r*\c) rectangle ({(\col+1)*\c}, {-(\r+1)*\c});
      \fi
    }
  }
 
  \fill[black!38] (1*\c, -3*\c) rectangle (2*\c, -4*\c);
  \fill[black!14] (2*\c, -3*\c) rectangle (3*\c, -4*\c);
  \fill[pattern=north east lines, pattern color=black!70] (2*\c, -3*\c) rectangle (3*\c, -4*\c);
  \draw[black!30, line width=0.25pt] (0,0) grid[step=\c] (6*\c, -6*\c);
  \draw[line width=0.4pt] (0,0) rectangle (6*\c, -6*\c);
  \draw[black!60, line width=0.4pt] (0,0) -- (6*\c, -6*\c);  % causal boundary (token level)
  \draw[line width=0.8pt] (0, -3*\c) rectangle (4*\c, -4*\c);
  \node[anchor=east, inner sep=1.5pt] at (0, -3.5*\c) {$Q_i$};
  \node[anchor=south, inner sep=1.5pt] at (1.5*\c, 0) {$K_j$};
  \node[anchor=north west, align=left, inner sep=0pt, font=\fontsize{7}{8.4}\selectfont]
        at (-0.1, -6*\c-0.08) {$Q_i$ visits $K_j$ for $j\le i$;\\
                               $J_{\mathrm{eff}}$ visits on average};
\end{scope}

\begin{scope}[shift={(1.62, -4.85)}, x=0.40cm, y=0.55cm]
  \newcommand{\tsact}[5]{% lane, start, end, style, label
    \draw[#4] (#2+0.03, #1+0.16) rectangle (#3-0.03, #1+0.84);
    \node[inner sep=0pt, font=\fontsize{7}{8.4}\selectfont] at ({(#2+#3)/2}, #1+0.5) {#5};}
  \node[lbl, anchor=west, inner xsep=0pt] at (-4.05, 4.7) {(c)};
  % legend
  \draw[act, fill=white] (-4.05, 3.85) rectangle (-3.45, 4.25);
  \node[anchor=west, inner sep=1.5pt] at (-3.45, 4.05) {visit $j$};
  \draw[act, fill=white, nxt] (-4.05, 3.2) rectangle (-3.45, 3.6);
  \node[anchor=west, inner sep=1.5pt] at (-3.45, 3.4) {visit $j{+}1$};
  % lanes
  \foreach \y/\name in {2/{SFU, CUDA}, 1/Tensor cores, 0/Memory} {
    \node[anchor=east, inner sep=1.5pt] at (-0.1, \y+0.5) {\name};
  }
  \foreach \y in {0,1,2,3} {
    \draw[black!22, line width=0.3pt] (0, \y) -- (13.4, \y);
    \draw[black!22, line width=0.3pt] (14.2, \y) -- (15.8, \y);
  }
  
  \begin{scope}[on background layer]
    \foreach \x in {1, 9, 13} \draw[guide] (\x, -0.1) -- (\x, 3.0);
  \end{scope}
  \foreach \y in {0,1,2} \node at (13.8, \y+0.5) {$\cdots$};
  % actions
  \tsact{0}{0}{1}{io}{$Q_i$}
  \tsact{0}{1}{2}{kk}{$K_j$}
  \tsact{0}{2}{3}{vv}{$V_j$}
  \tsact{0}{3}{4}{kk,nxt}{}
  \tsact{0}{4}{5}{vv,nxt}{}
  \tsact{1}{2}{3.5}{tc}{$QK^{\!\top}$}
  \tsact{1}{4}{5.5}{tc,nxt}{}
  \tsact{2}{3.5}{7.5}{sm}{softmax$_j$}
  \tsact{1}{7.5}{9}{tc}{$PV$}
  \tsact{2}{7.5}{11.5}{sm,nxt}{}
  \node[fill=green!28, inner xsep=1pt, inner ysep=0.5pt, font=\fontsize{7}{8.4}\selectfont] at (9.5, 2.5) {softmax$_{j+1}$};
  \tsact{1}{11.5}{13}{tc,nxt}{}
  \tsact{0}{14.5}{15.5}{io}{$O_i$}
  % PV_j waits for softmax_j: tensor cores idle
  \node[font=\fontsize{7}{8.4}\selectfont, text=black!55] at (6.5, 1.5) {idle};
  \draw[dep, densely dotted] (7.1, 2.16) to[out=-90, in=90] (7.9, 1.84);
  % equation terms
  \draw[guide] (1, 3.0) -- (1, 4.25);  \draw[guide] (13, 3.0) -- (13, 4.25);
  \draw[guide] (9, 3.0) -- (9, 3.25);
  \draw[br] (1, 3.25) -- node[above=4.5pt, inner sep=0pt] {$T_1$: one visit} (9, 3.25);
  \draw[br] (1, 4.25) -- node[above=4.5pt, inner sep=0pt] {$T_2$: two consecutive visits} (13, 4.25);
  \draw[br, decoration={mirror}] (0, -0.12) -- node[below=4.5pt, inner sep=0pt] {$T_q$} (1, -0.12);
  \draw[br, decoration={mirror}] (9, -0.12) -- node[below=4.5pt, inner sep=0pt, align=center]
        {$T_2-T_1$: added by\\each further visit} (13, -0.12);
  \draw[br, decoration={mirror}] (14.5, -0.12) -- node[below=4.5pt, inner sep=0pt] {$T_o$} (15.5, -0.12);
\end{scope}
\end{tikzpicture}
\caption{Prefill attention for one query block, showing the sequence of operations. (a) In its update with key block $j$, query block $i$ loads $K_j$ and $V_j$, computes $Q_iK_j^\top$, applies softmax, and multiplies the result by $V_j$. The next update uses the updated running maximum and adds to the same output $O_i$. (b) With causal masking, query block $i$ processes only key blocks $j\le i$. (c) Consecutive updates overlap: while softmax for update $j$ runs, the tensor cores begin processing update $j+1$. Thus, two updates take $T_2$ rather than $2T_1$, and each subsequent update adds $T_2-T_1$, which here equals the softmax time, the busiest resource per update ($u_{\max}$).}
  \label{fig:attn}
   \vspace{-15pt}
\end{figure}
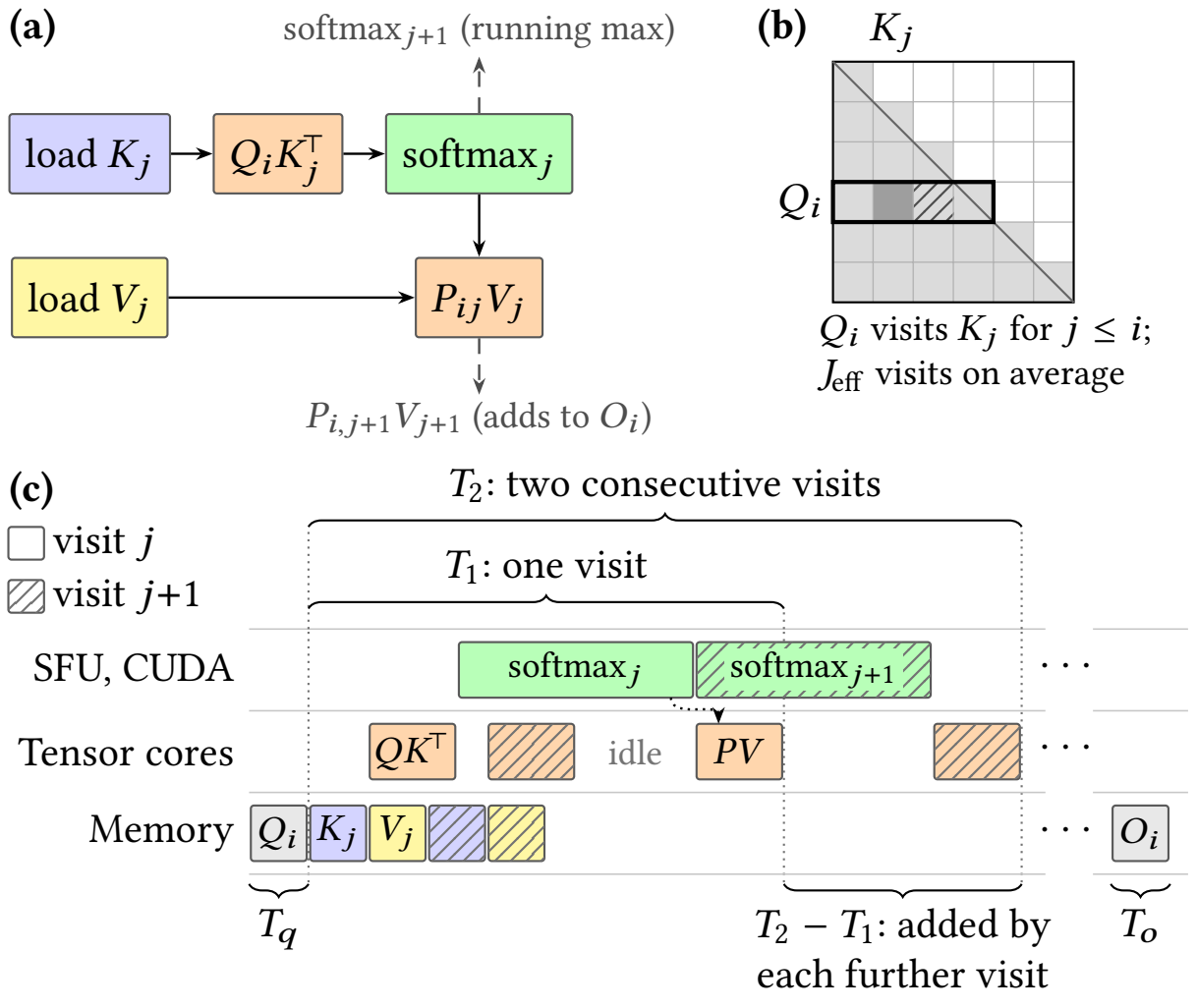

%% file: tab_const.tex
\begin{table}[t]
  \centering
  \footnotesize
  \setlength{\tabcolsep}{2pt}
  \caption{Main hardware constants. $^\dagger$dense, flop/clk/SM.}
   \vspace{-10pt}
  \label{tab:const}
  \begin{tabular}{@{}lll@{}}
    \toprule
    constant & H100 & GB300 \\
    \midrule
    SMs & 132 & 152 \\
    SM clock, MHz & 1,830 & 1,957; 1,335--1,447 \\
    FP8; NVFP4 rate$^\dagger$ & 5,327; -- & 15,971; 29,061 \\
    HBM bandwidth & 3.13 TB/s & 7.4 TB/s \\
    L2 bandwidth; size & 7.18 TB/s; 50 MB & 10.2 TB/s; 128 MiB \\
    L1.5 bandwidth; size & -- & 22.5 TB/s; 2 MB$^\ddagger$ \\
    TMEM read; write per SM & -- & 2.08; 1.47 TB/s \\
    NVLink BW $(\beta_{\text{NV}})$ & 396 GB/s & 587--784 GB/s \\
    collective latency $(\alpha)$ & 6.6--19.9; 5 $\mu$s & 3.1--49.8 $\mu$s \\
    host time per step & 6.0--52 $\mu$s, added & 1.5--10.9 ms, floor \\
    GPU memory & 81,081 MiB & 276.5 GiB \\
    \bottomrule
  \end{tabular}
   \vspace{-15pt}
\end{table}

%% file: fig_parity_all.tex
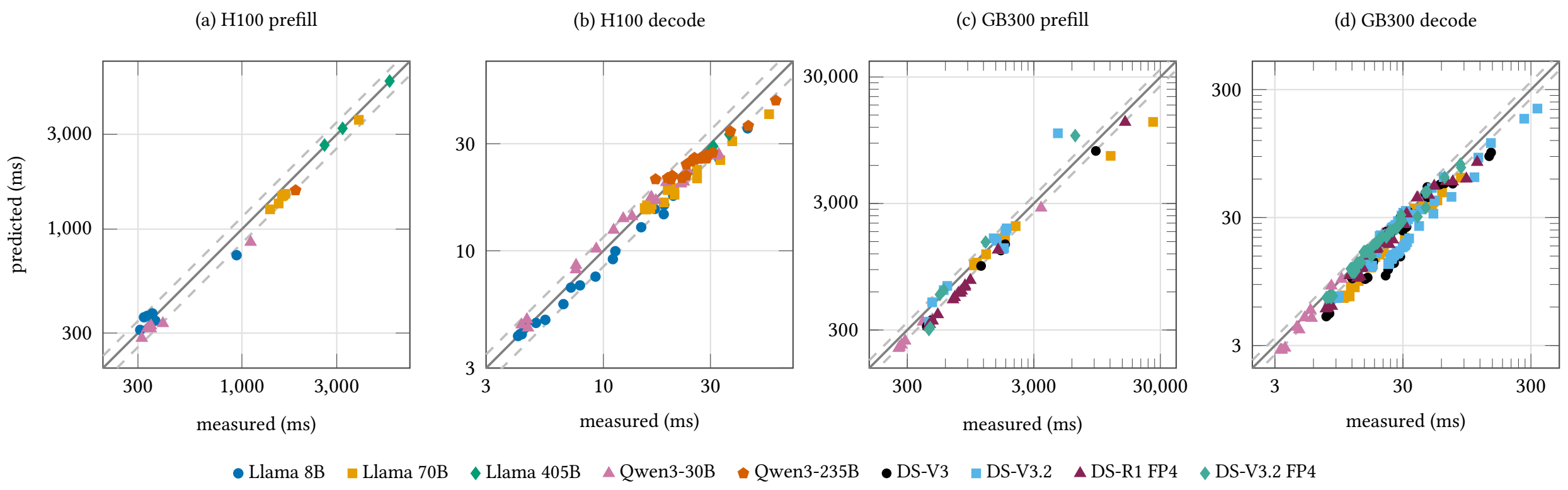
\begin{figure*}[t]
\centering
\begin{tikzpicture}
\begin{groupplot}[osplit, group style={group size=4 by 1, horizontal sep=0.8cm},
  width=0.27\textwidth, height=0.27\textwidth, xmode=log, ymode=log, xlabel={measured (ms)}, log ticks with fixed point]
\nextgroupplot[legend to name=parityleg, legend columns=9, legend style={/tikz/every even column/.append style={column sep=5pt}}, ylabel={predicted (ms)}, xtick={300,1000,3000}, ytick={300,1000,3000}, title={(a) H100 prefill}, xmin=200, xmax=7000, ymin=200, ymax=7000]
\addlegendimage{only marks, m8b}\addlegendentry{Llama 8B}
\addlegendimage{only marks, m70b}\addlegendentry{Llama 70B}
\addlegendimage{only marks, m405b}\addlegendentry{Llama 405B}
\addlegendimage{only marks, mq30}\addlegendentry{Qwen3-30B}
\addlegendimage{only marks, mq235}\addlegendentry{Qwen3-235B}
\addlegendimage{only marks, mdv3}\addlegendentry{DS-V3}
\addlegendimage{only marks, mdv32}\addlegendentry{DS-V3.2}
\addlegendimage{only marks, mdr1}\addlegendentry{DS-R1 FP4}
\addlegendimage{only marks, mdv32f}\addlegendentry{DS-V3.2 FP4}
\addplot[black!50, mark=none, domain=200:7000, samples=2, forget plot] {1*x};
\addplot[black!25, dashed, mark=none, domain=200:7000, samples=2, forget plot] {1.15*x};
\addplot[black!25, dashed, mark=none, domain=200:7000, samples=2, forget plot] {0.85*x};
\addplot[forget plot, only marks, m8b] coordinates {(321.6,360.2) (366.8,346.4) (333.4,365.8) (940.2,739.5) (354.6,376.8) (307.9,311.2)};
\addplot[forget plot, only marks, m70b] coordinates {(1590,1463) (1531,1346) (1617,1477) (3875,3550) (1655,1505) (1386,1258)};
\addplot[forget plot, only marks, m405b] coordinates {(3211,3215) (5536,5558) (2605,2649)};
\addplot[forget plot, only marks, mq30] coordinates {(334.7,317.6) (400.5,335.5) (349.1,326.8) (1106,860.3) (348.3,315.7) (313.6,282.8)};
\addplot[forget plot, only marks, mq235] coordinates {(1864,1569)};
\nextgroupplot[xtick={3,10,30}, ytick={3,10,30}, title={(b) H100 decode}, xmin=3, xmax=70, ymin=3, ymax=70]
\addplot[black!50, mark=none, domain=3:70, samples=2, forget plot] {1*x};
\addplot[black!25, dashed, mark=none, domain=3:70, samples=2, forget plot] {1.15*x};
\addplot[black!25, dashed, mark=none, domain=3:70, samples=2, forget plot] {0.85*x};
\addplot[forget plot, only marks, m8b] coordinates {(4.164,4.168) (9.221,7.669) (14.74,12.73) (5.508,4.922) (25.92,22.46) (6.637,5.783) (4.329,4.246) (4.456,4.519) (16.92,15.36) (7.167,6.854) (4.26,4.258) (18.67,16.02) (7.867,7.009) (11.3,9.956) (5.009,4.768) (11.02,9.177) (18.57,14.55) (32.82,25.75) (43.93,35.13) (20.48,17.59)};
\addplot[forget plot, only marks, m70b] coordinates {(15.55,15.27) (26.15,21.01) (37.53,30.77) (18.62,16.43) (20.75,17.77) (16.19,15.38) (15.64,15.95) (19.34,18.64) (15.28,15.51) (20.7,19.04) (26.14,22.99) (16.41,16.03) (33.19,25.37) (54.76,40.64)};
\addplot[forget plot, only marks, m405b] coordinates {(36.35,33.02) (30,27.93) (30.91,29.22)};
\addplot[forget plot, only marks, mq30] coordinates {(4.616,4.534) (17.13,16.77) (22.18,19.9) (11.1,12.37) (31.83,26.47) (13.42,14.23) (7.546,8.239) (4.562,4.942) (19.06,20.19) (9.293,10.2) (4.318,4.68) (23.57,23.03) (12.26,13.93) (16.36,17.36) (7.552,8.631) (22.82,20.28) (32.64,26.95)};
\addplot[forget plot, only marks, mq235] coordinates {(25.41,25.92) (30.71,27.35) (27.18,25.96) (17.08,20.84) (28.67,26.62) (28.67,25.77) (28.91,25.97) (28.8,25.97) (24.52,25.02) (28.79,25.97) (28.9,25.97) (26.75,25.7) (20,20.84) (20.25,21.63) (23.43,24.25) (23.37,21.63) (19.34,21.12) (25.54,25.85) (25.44,25.44) (25.54,25.85) (22.4,21.12) (58.67,46.74) (36.84,34.04) (44.25,36.05)};
\nextgroupplot[xtick={300,3000,30000}, ytick={300,3000,30000}, title={(c) GB300 prefill}, xmin=150, xmax=40000, ymin=150, ymax=40000]
\addplot[black!50, mark=none, domain=150:40000, samples=2, forget plot] {1*x};
\addplot[black!25, dashed, mark=none, domain=150:40000, samples=2, forget plot] {1.15*x};
\addplot[black!25, dashed, mark=none, domain=150:40000, samples=2, forget plot] {0.85*x};
\addplot[forget plot, only marks, m70b] coordinates {(1654,1547) (1674,1562) (1709,1592) (1777,1650) (2155,1988) (2.629e+04,1.324e+04) (1010,971.7) (1017,979.2) (1034,994) (1065,1023) (1258,1192) (1.214e+04,7138)};
\addplot[forget plot, only marks, mq30] coordinates {(256.3,215.3) (260.8,219.8) (270,228.7) (288.4,246.1) (398.6,347.5) (3422,2771)};
\addplot[forget plot, only marks, mdv3] coordinates {(1647,1276) (1659,1298) (1698,1341) (1780,1428) (427.6,319.4) (432.6,325) (443.9,335.8) (464.9,357.4) (1149,962.7) (9273,7807)};
\addplot[forget plot, only marks, mdv32] coordinates {(1705,1310) (1746,1331) (1527,1521) (1792,1828) (1828,1908) (446.1,343.2) (450.8,348.8) (471.8,496.5) (582.1,620) (620.8,669.2) (1445,1592) (4647,1.077e+04)};
\addplot[forget plot, only marks, mdr1] coordinates {(689.9,521.5) (710.4,543.3) (761.4,587.1) (857.4,673.7) (787,590.2) (806.5,612) (854,655.8) (944.8,742.4) (439.9,321.6) (451.4,332.6) (475.6,354.4) (522.4,397.8) (1565,1290) (1.58e+04,1.319e+04)};
\addplot[forget plot, only marks, mdv32f] coordinates {(444.6,302.7) (454.2,313.7) (541.9,571.9) (581.2,621.1) (1248,1485) (6375,1.032e+04)};
\nextgroupplot[xtick={3,30,300}, ytick={3,30,300}, title={(d) GB300 decode}, xmin=2, xmax=500, ymin=2, ymax=500]
\addplot[black!50, mark=none, domain=2:500, samples=2, forget plot] {1*x};
\addplot[black!25, dashed, mark=none, domain=2:500, samples=2, forget plot] {1.15*x};
\addplot[black!25, dashed, mark=none, domain=2:500, samples=2, forget plot] {0.85*x};
\addplot[forget plot, only marks, m70b] coordinates {(16.55,12.22) (16.95,12.22) (17.45,12.86) (21.34,16.16) (50.57,35.83) (85.14,61.44) (16.75,12.27) (17.54,12.87) (19.76,15.44) (30.51,25.19) (51.66,39.43) (17.06,12.59) (19.82,15.45) (28.94,25.77) (43.35,39.89) (10.26,7.085) (10.29,7.082) (11.51,7.262) (13.54,9.579) (31.01,20.84) (55.78,40.83) (10.4,7.089) (10.59,7.394) (12.63,8.552) (18.3,14.74) (48.98,38.89) (60.7,47.15) (10.76,7.252) (11.76,8.684) (17.24,13.72) (36.65,35.4) (45.52,42.6)};
\addplot[forget plot, only marks, mq30] coordinates {(3.317,2.771) (4.651,3.993) (5.809,4.91) (8.051,6.825) (14.59,14.52) (22.64,22.17) (3.398,2.774) (4.454,4.18) (5.689,5.683) (9.974,9.919) (24.99,25.35) (43.96,43.06) (3.602,2.872) (5.122,4.954) (8.285,8.783) (20.79,22.32) (39.58,42.76)};
\addplot[forget plot, only marks, mdv3] coordinates {(21.98,10.55) (23,11.76) (23.7,12.67) (25.6,13.21) (28.64,14.86) (15.21,9.91) (16.69,13.35) (19.35,18.93) (23.43,21.92) (29.88,23.77) (146.6,96.74) (15.34,9.987) (16.78,13.43) (18.95,19.01) (23.98,22.32) (32.04,25.37) (141.6,90.4) (15.92,10.19) (17.64,13.63) (19.23,19.34) (25.08,23.93) (37.76,31.79) (73.48,55.19) (7.554,5.058) (12.07,10.04) (19.08,19.27) (28.81,30.3) (46.77,43.25) (63.62,55.59) (7.666,5.135) (11.88,10.25) (19.9,20.07) (32.86,33.51) (58.58,54.47) (62.39,57.86) (8.007,5.335) (12.85,10.98) (22.32,23.28) (44.8,46.33) (46.56,51.94)};
\addplot[forget plot, only marks, mdv32] coordinates {(23.02,13.01) (23.98,14.22) (24.93,15.14) (26.63,15.64) (29.84,17.56) (337.2,213.2) (23.05,13.09) (24.04,14.3) (25.01,15.22) (26.88,15.81) (30.46,17.9) (266.1,177.6) (23.26,13.03) (24.31,14.25) (25.27,15.16) (27.26,15.8) (31.13,18.17) (108.6,61.98) (23.26,13.05) (24.47,14.26) (25.33,15.17) (27.38,15.92) (31.52,18.53) (71.88,43.53) (23.34,13.11) (24.43,14.33) (25.45,15.24) (27.6,16.09) (32.36,19.25) (51.92,32.17) (23.52,13.27) (24.76,14.48) (25.74,15.4) (28.22,16.45) (33.55,20.6) (40.17,25.76) (16.43,12.37) (17.55,15.81) (20.09,21.36) (24.41,24.62) (31.62,27.77) (146.1,114.7) (16.53,12.45) (17.59,15.9) (19.74,21.53) (24.79,24.96) (32.91,29.11) (116.5,88.21) (16.97,12.4) (18.3,15.84) (20.53,21.52) (24.98,25.23) (34.44,30.3) (52.86,40.47) (16.99,12.41) (18.16,15.85) (20.6,21.64) (25.72,25.59) (36.28,31.73) (41.58,34.71) (17.08,12.48) (18.45,15.92) (20.12,21.82) (26.27,26.31) (35.73,31.5) (17.25,12.63) (18.57,16.08) (22.17,22.17) (29.18,27.66) (31.61,29.19) (8.648,7.04) (12.73,11.9) (20.3,20.15) (29.27,31.04) (46.7,47.34) (62.41,62.01) (8.763,7.107) (13.37,12.01) (20.14,20.57) (30.17,32.74) (50.36,51.09) (9.137,7.052) (13.48,12.13) (21.06,21.06) (31.64,33.82) (9.209,7.064) (13.61,12.3) (21.9,21.73) (27.46,28.16) (9.299,7.132) (14.02,12.66) (23.42,22.53) (9.486,7.288) (15.42,13.33) (20.71,18.78)};
\addplot[forget plot, only marks, mdr1] coordinates {(13.2,10.1) (14.02,12.03) (15.74,15.03) (18.61,16.93) (23.45,18.54) (93.91,60.25) (13.38,10.17) (14.32,12.11) (16.19,15.11) (18.95,17.33) (24.98,20.15) (114.1,81.21) (13.95,10.37) (14.4,12.31) (16.14,15.44) (20.08,18.93) (30.87,26.57) (73.49,57.21) (7.508,5.821) (11.54,10.6) (19.43,18.95) (29.38,28.93) (47.79,44.99) (65.38,57.54) (7.781,5.897) (12.52,10.81) (19.97,19.75) (32.08,32.14) (52.69,52.9) (8.347,6.098) (12.71,11.54) (22.88,22.96) (38.74,42.97)};
\addplot[forget plot, only marks, mdv32f] coordinates {(12.04,11.21) (13.14,13.15) (14.77,16.11) (17.59,18.22) (22.33,21.17) (85.38,73.81) (12.11,11.29) (13.23,13.23) (15.07,16.29) (17.99,18.55) (23.37,22.51) (84.32,78.27) (12.46,11.24) (13.59,13.17) (15.47,16.28) (18.42,18.83) (24.9,23.7) (45.25,35.85) (12.55,11.25) (13.69,13.19) (15.71,16.39) (18.96,19.19) (26.59,25.13) (39.08,30.47) (12.65,11.32) (13.68,13.25) (15.86,16.57) (19.77,19.9) (29.15,27.52) (12.8,11.48) (14.1,13.41) (16.63,16.93) (22.2,21.25) (26.88,24.75) (7.588,7.114) (11.72,11.94) (19.2,19.64) (28.62,29.7) (46.56,47.87) (63.51,61.83) (7.837,7.181) (12.02,12.05) (19.61,20.06) (29.1,31.4) (44.99,47.27) (8.21,7.125) (12.55,12.17) (19.68,20.55) (29.61,28.1) (8.274,7.138) (12.78,12.34) (20.75,21.23) (23.94,23.73) (8.374,7.206) (13.38,12.7) (18.97,19.01) (8.527,7.362) (14.18,13.37) (17.06,15.01)};
\end{groupplot}
\end{tikzpicture}\\[1pt]\pgfplotslegendfromname{parityleg}
\caption{Predicted against measured step time, every scored point: (a, b) H100; (c, d) GB300. Solid line: equality; dashed: $\pm$15\%. DS: DeepSeek.}
\label{fig:parity}
 \vspace{-10pt}
\end{figure*}

%% file: tab_acc.tex
\begin{table}[t]
  \centering
  \footnotesize
  \setlength{\tabcolsep}{3.5pt}
  \caption{MAPE (\%) of OSplit and of the roofline, with the number of points in parentheses. DS: DeepSeek.}
  \label{tab:acc}
   \vspace{-10pt}
  \begin{tabular}{@{}lrrrrr@{}}
    \toprule
     & KV & \multicolumn{2}{c}{prefill} & \multicolumn{2}{c}{decode} \\
    \cmidrule(lr){3-4}\cmidrule(l){5-6}
     & capacity & OSplit & roofline & OSplit & roofline \\
    \midrule
    \multicolumn{6}{@{}l}{\emph{\textbf{H100}}} \\
    Llama 8B & 0.9\,(2) & 9.3\,(6) & 55.0\,(6) & 11.0\,(20) & 40.7\,(15) \\
    Llama 70B & 2.6\,(1) & 9.2\,(6) & 48.2\,(6) & 10.7\,(14) & 35.2\,(12) \\
    Llama 405B & 6.7\,(1) & 0.7\,(3) & 43.3\,(3) & 7.2\,(3) & 47.2\,(3) \\
    Qwen3-30B & 3.3\,(1) & 11.5\,(6) & 75.4\,(6) & 9.1\,(17) & 40.7\,(15) \\
    Qwen3-235B & 4.0\,(6) & 15.8\,(1) & -- & 7.9\,(24) & -- \\
    all & 3.5\,(11) & 9.0\,(22) & 57.2\,(21) & 9.4\,(78) & 39.7\,(45) \\
    \midrule
    \multicolumn{6}{@{}l}{\emph{\textbf{GB300}}} \\
    Llama 70B & 0.4\,(2) & 12.2\,(12) & 47.5\,(12) & 24.3\,(32) & 45.1\,(32) \\
    Qwen3-30B & 1.2\,(1) & 15.6\,(6) & 75.0\,(6) & 8.1\,(17) & 45.8\,(17) \\
    DS-V3 & 6.5\,(3) & 21.5\,(10) & 63.8\,(10) & 19.8\,(40) & 56.6\,(40) \\
    DS-V3.2 & 7.1\,(2) & 21.7\,(12) & 65.8\,(12) & 22.4\,(95) & 66.2\,(95) \\
    DS-R1 FP4 & 5.5\,(2) & 23.0\,(14) & 47.4\,(14) & 13.2\,(33) & 46.9\,(33) \\
    DS-V3.2 FP4 & 6.4\,(2) & 26.0\,(6) & 58.5\,(6) & 6.0\,(59) & 52.4\,(59) \\
    all & 5.0\,(12) & 19.9\,(60) & 57.7\,(60) & 16.8\,(276) & 55.8\,(276) \\
    \bottomrule
  \end{tabular}
  \vspace{-15pt}
\end{table}

%% file: tab_regret.tex
\begin{table}[t]
  \centering
  \footnotesize
  \setlength{\tabcolsep}{4pt}
  \caption{The split on one 8-GPU H100 node: the best number of prefill GPUs, the same for the measurement and both methods, and the median error of each method's throughput curve.}
   \vspace{-10pt}
  \label{tab:regret}
  \begin{tabular}{@{}llcrr@{}}
    \toprule
     & & best $m$ & \multicolumn{2}{c}{curve error} \\
     \cmidrule(l){4-5}
     & & (all three) & OSplit & AIConfigurator \\
    \midrule
    Llama 8B & W1 & 1 & 16.3\% & 5.0\% \\
     & W2 & 4 & 13.1\% & 14.1\% \\
     & W3 & 6 & 25.0\% & 20.9\% \\
    \midrule
    Qwen3-30B & W1 & 1 & 3.7\% & 38.1\% \\
     & W2 & 3 & 8.4\% & 30.0\% \\
     & W3 & 6 & 27.0\% & 19.5\% \\
    \bottomrule
  \end{tabular}
   \vspace{-10pt}
\end{table}

%% file: fig_split_C.tex
\begin{figure}[t]
\centering
\begin{tikzpicture}
\begin{groupplot}[osplit, group style={group size=2 by 3, horizontal sep=22pt, vertical sep=6pt, x descriptions at=edge bottom},
 width=0.585\columnwidth, height=2.7cm, xmin=0.5, xmax=7.5, xtick={1,...,7}, ymin=0]
\nextgroupplot[legend to name=altC, legend columns=3, title={Llama 8B}, title style={yshift=-1.5ex}, ylabel={W1, req/s}]
\addplot[black!45, dashed, mark=none] coordinates {(1,36.5) (2,31.3) (3,26.1) (4,20.8) (5,15.6) (6,10.4) (7,5.21)};
\addplot[c8b, mark=none, line width=1pt] coordinates {(1,44.7) (2,38.3) (3,31.9) (4,25.5) (5,19.1) (6,12.8) (7,6.38)};
\addplot[only marks, mark=*, mark size=1.2pt, black] coordinates {(1,38.4) (2,32.9) (3,27.5) (4,22.1) (5,16.5) (6,10.9) (7,5.47)};
\legend{AIConfigurator, OSplit, measured}
\nextgroupplot[title={Qwen3-30B}, title style={yshift=-1.5ex}]
\addplot[black!45, dashed, mark=none] coordinates {(1,20.1) (2,17.2) (3,14.4) (4,11.5) (5,8.62) (6,5.75) (7,2.87)};
\addplot[c8b, mark=none, line width=1pt] coordinates {(1,33.7) (2,28.9) (3,24.1) (4,19.2) (5,14.4) (6,9.62) (7,4.81)};
\addplot[only marks, mark=*, mark size=1.2pt, black] coordinates {(1,32.5) (2,28) (3,23.3) (4,18.3) (5,13.9) (6,9.2) (7,4.61)};
\nextgroupplot[ylabel={W2, req/s}]
\addplot[black!45, dashed, mark=none] coordinates {(1,5.57) (2,11.1) (3,16.7) (4,19.5) (5,14.6) (6,9.73) (7,4.87)};
\addplot[c8b, mark=none, line width=1pt] coordinates {(1,7.57) (2,15.1) (3,22.7) (4,24.9) (5,18.6) (6,12.4) (7,6.21)};
\addplot[only marks, mark=*, mark size=1.2pt, black] coordinates {(1,6.74) (2,13.5) (3,20.4) (4,21.8) (5,16.4) (6,10.9)};
\nextgroupplot[]
\addplot[black!45, dashed, mark=none] coordinates {(1,5.68) (2,11.4) (3,15.1) (4,11.5) (5,8.66) (6,5.77) (7,2.89)};
\addplot[c8b, mark=none, line width=1pt] coordinates {(1,7.36) (2,14.7) (3,22.1) (4,17.2) (5,12.9) (6,8.62) (7,4.31)};
\addplot[only marks, mark=*, mark size=1.2pt, black] coordinates {(1,6.45) (2,12.9) (3,19.4) (4,18) (5,14.1) (6,8.62) (7,4.12)};
\nextgroupplot[ylabel={W3, req/s}, xlabel={prefill GPUs $m$}]
\addplot[black!45, dashed, mark=none] coordinates {(1,0.856) (2,1.71) (3,2.57) (4,3.42) (5,4.28) (6,5.13) (7,3.65)};
\addplot[c8b, mark=none, line width=1pt] coordinates {(1,1.35) (2,2.7) (3,4.06) (4,5.41) (5,6.76) (6,8.11) (7,4.63)};
\addplot[only marks, mark=*, mark size=1.2pt, black] coordinates {(1,1.07) (2,2.13) (3,3.25) (4,4.33) (5,5.44) (6,6.5) (7,3.54)};
\nextgroupplot[xlabel={prefill GPUs $m$}]
\addplot[black!45, dashed, mark=none] coordinates {(1,0.736) (2,1.47) (3,2.21) (4,2.94) (5,3.68) (6,4.42) (7,2.56)};
\addplot[c8b, mark=none, line width=1pt] coordinates {(1,1.16) (2,2.33) (3,3.49) (4,4.65) (5,5.81) (6,6.92) (7,3.59)};
\addplot[only marks, mark=*, mark size=1.2pt, black] coordinates {(1,0.907) (2,1.82) (3,2.75) (4,3.67) (5,4.57) (6,5.49) (7,3.64)};
\end{groupplot}\end{tikzpicture}\\[1pt]\pgfplotslegendfromname{altC}
\caption{Throughput of every split of one 8-GPU H100 node into $m$ prefill and $8-m$ decode GPUs, one GPU per instance. Both methods choose the measured best split in all six cases.}
\label{fig:split}
\vspace{-15pt}
\end{figure}
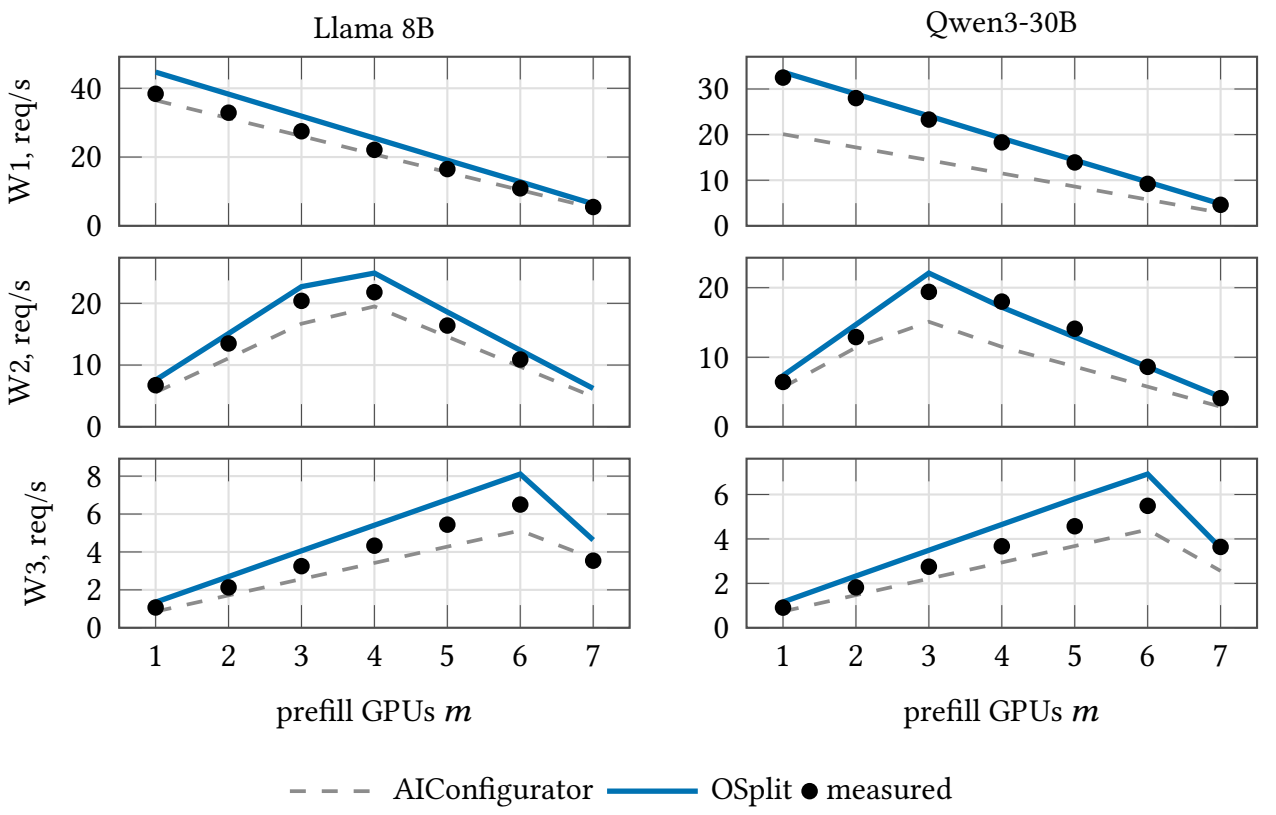

%% file: fig_ratio.tex
\begin{figure}[t]
\centering
\begin{tikzpicture}
\begin{groupplot}[group style={group size=2 by 1, horizontal sep=14pt, y descriptions at=edge left},
  osplit, width=0.57\columnwidth, height=4.6cm, ymode=log, ymin=0.08, ymax=80, ytick={0.1,1,10}, yticklabels={0.1,1,10}, xmode=log, log basis x=2, xtick={1024,4096,16384,65536}, xticklabels={1K,4K,16K,64K}, xlabel={input length $I$}]
\nextgroupplot[xmin=1024, xmax=65536, ylabel={best ratio $r^*$ (prefill/decode GPUs)}, title={(a) H100, $N=8$}, title style={yshift=-4pt}, legend to name=ratlegA, legend columns=3]
\addplot[fill=c8b, fill opacity=0.07, draw=none, forget plot] coordinates {(1024,0.1226) (1536,0.1597) (2048,0.1862) (2816,0.2165) (4096,0.2533) (5888,0.2904) (8192,0.3136) (11520,0.3021) (16384,0.3845) (23296,0.4374) (32768,0.4428) (46336,0.5027) (65536,0.6298) (65536,5.104) (46336,4.088) (32768,3.635) (23296,3.698) (16384,3.327) (11520,2.569) (8192,2.738) (5888,2.622) (4096,2.45) (2816,2.175) (2048,2.012) (1536,1.83) (1024,1.508)} -- cycle;
\addplot[m8b] coordinates {(1024,0.5141) (1536,0.605) (2048,0.6827) (2816,0.7777) (4096,0.8745) (5888,0.9537) (8192,1.019) (11520,0.952) (16384,1.221) (23296,1.378) (32768,1.36) (46336,1.533) (65536,1.915)};
\addlegendentry{Llama 8B}
\addplot[fill=c70b, fill opacity=0.07, draw=none, forget plot] coordinates {(1024,0.1972) (1536,0.2658) (2048,0.3336) (2816,0.4234) (4096,0.4827) (5888,0.5582) (8192,0.6871) (11520,0.7677) (16384,0.8198) (23296,1.026) (32768,1.149) (46336,1.432) (65536,1.29) (65536,10.41) (46336,11.6) (32768,9.466) (23296,8.488) (16384,6.759) (11520,6.746) (8192,5.806) (5888,5.272) (4096,4.764) (2816,3.91) (2048,3.356) (1536,3.011) (1024,2.313)} -- cycle;
\addplot[m70b] coordinates {(1024,0.8053) (1536,0.9869) (2048,1.171) (2816,1.425) (4096,1.751) (5888,1.942) (8192,2.155) (11520,2.472) (16384,2.525) (23296,3.176) (32768,3.511) (46336,4.352) (65536,3.909)};
\addlegendentry{Llama 70B}
\addplot[fill=c405b, fill opacity=0.07, draw=none, forget plot] coordinates {(1024,0.2446) (1536,0.321) (2048,0.3799) (2816,0.4448) (4096,0.5767) (5888,0.6258) (8192,0.7337) (11520,0.8258) (16384,0.8593) (23296,0.9678) (32768,0.9299) (46336,1.001) (46336,8.024) (32768,8.023) (23296,8) (16384,7.717) (11520,7.015) (8192,6.697) (5888,6.608) (4096,5.916) (2816,5.432) (2048,4.161) (1536,3.573) (1024,2.638)} -- cycle;
\addplot[m405b] coordinates {(1024,0.9135) (1536,1.2) (2048,1.508) (2816,1.871) (4096,2.003) (5888,2.414) (8192,2.364) (11520,2.56) (16384,2.801) (23296,2.925) (32768,3.018) (46336,3.019)};
\addlegendentry{Llama 405B}
\addplot[fill=cq30, fill opacity=0.07, draw=none, forget plot] coordinates {(1024,0.09703) (1536,0.125) (2048,0.1522) (2816,0.1821) (4096,0.2213) (5888,0.2657) (8192,0.2951) (11520,0.3638) (16384,0.4051) (23296,0.4298) (32768,0.55) (32768,4.483) (23296,3.829) (16384,3.366) (11520,3.142) (8192,2.668) (5888,2.486) (4096,2.17) (2816,1.905) (2048,1.71) (1536,1.513) (1024,1.278)} -- cycle;
\addplot[mq30] coordinates {(1024,0.365) (1536,0.4706) (2048,0.5249) (2816,0.6029) (4096,0.6931) (5888,0.8143) (8192,0.8965) (11520,1.061) (16384,1.142) (23296,1.272) (32768,1.528)};
\addlegendentry{Qwen3-30B}
\addplot[fill=cq235, fill opacity=0.07, draw=none, forget plot] coordinates {(1024,0.119) (1536,0.1656) (2048,0.1972) (2816,0.2296) (4096,0.3005) (5888,0.3395) (8192,0.3887) (11520,0.4713) (16384,0.4636) (23296,0.5406) (32768,0.6673) (32768,5.553) (23296,5.005) (16384,4.241) (11520,4.396) (8192,3.556) (5888,3.558) (4096,3.048) (2816,2.657) (2048,2.328) (1536,1.988) (1024,1.519)} -- cycle;
\addplot[mq235] coordinates {(1024,0.4635) (1536,0.6044) (2048,0.71) (2816,0.811) (4096,0.9338) (5888,1.139) (8192,1.144) (11520,1.453) (16384,1.414) (23296,1.674) (32768,1.898)};
\addlegendentry{Qwen3-235B}
\nextgroupplot[xmin=1024, xmax=32768, title={(b) GB300, $N=72$}, title style={yshift=-4pt}, legend to name=ratlegB, legend columns=4]
\addplot[fill=c70b, fill opacity=0.07, draw=none, forget plot] coordinates {(1024,0.2462) (2048,0.3865) (4096,0.5497) (8192,0.7323) (32768,0.8759) (32768,7.16) (8192,6.262) (4096,5.212) (2048,4.063) (1024,2.722)} -- cycle;
\addplot[m70b, forget plot] coordinates {(1024,0.9155) (2048,1.334) (4096,1.859) (8192,2.229) (32768,2.601)};
\addplot[fill=cq30, fill opacity=0.07, draw=none, forget plot] coordinates {(1024,0.1216) (2048,0.199) (4096,0.2806) (8192,0.3622) (32768,0.5466) (32768,4.711) (8192,3.246) (4096,2.712) (2048,1.927) (1024,1.242)} -- cycle;
\addplot[mq30, forget plot] coordinates {(1024,0.4186) (2048,0.662) (4096,0.9539) (8192,1.154) (32768,1.687)};
\addplot[fill=black, fill opacity=0.07, draw=none, forget plot] coordinates {(1024,0.386) (2048,0.7822) (4096,1.321) (8192,2.08) (32768,3.923) (32768,33.76) (8192,17.87) (4096,11.98) (2048,6.356) (1024,3.45)} -- cycle;
\addplot[mdv3] coordinates {(1024,1.237) (2048,2.236) (4096,4.151) (8192,6.563) (32768,12.1)};
\addlegendentry{DS-V3}
\addplot[fill=cdv, fill opacity=0.07, draw=none, forget plot] coordinates {(1024,0.3405) (2048,0.617) (4096,1.631) (8192,2.694) (32768,9.055) (32768,74.75) (8192,23.81) (4096,13.4) (2048,5.03) (1024,2.965)} -- cycle;
\addplot[mdv32] coordinates {(1024,1.008) (2048,1.835) (4096,4.864) (8192,8.458) (32768,27.59)};
\addlegendentry{DS-V3.2}
\addplot[fill=cr1, fill opacity=0.07, draw=none, forget plot] coordinates {(1024,0.2544) (2048,0.457) (4096,0.8165) (8192,1.205) (32768,3.173) (32768,27.25) (8192,11.09) (4096,7.097) (2048,4.179) (1024,2.267)} -- cycle;
\addplot[mdr1] coordinates {(1024,0.8028) (2048,1.486) (4096,2.657) (8192,3.993) (32768,9.81)};
\addlegendentry{DS-R1 FP4}
\addplot[fill=cv32f, fill opacity=0.07, draw=none, forget plot] coordinates {(1024,0.1613) (2048,0.3277) (4096,1.186) (8192,2.119) (32768,5.508) (32768,45.91) (8192,18.77) (4096,10.56) (2048,2.657) (1024,1.581)} -- cycle;
\addplot[mdv32f] coordinates {(1024,0.5407) (2048,0.9683) (4096,3.745) (8192,6.721) (32768,16.55)};
\addlegendentry{DS-V3.2 FP4}
\end{groupplot}
\end{tikzpicture}\\[1pt]
\pgfplotslegendfromname{ratlegA}\\\pgfplotslegendfromname{ratlegB}
\caption{Best ratio $r^*$ of prefill to decode GPUs against input length. Lines: mean output length of about 700 tokens; shaded: 2,048 (bottom) to 256 (top). DS: DeepSeek.}
\vspace{-15pt}
\label{fig:ratio}
\end{figure}
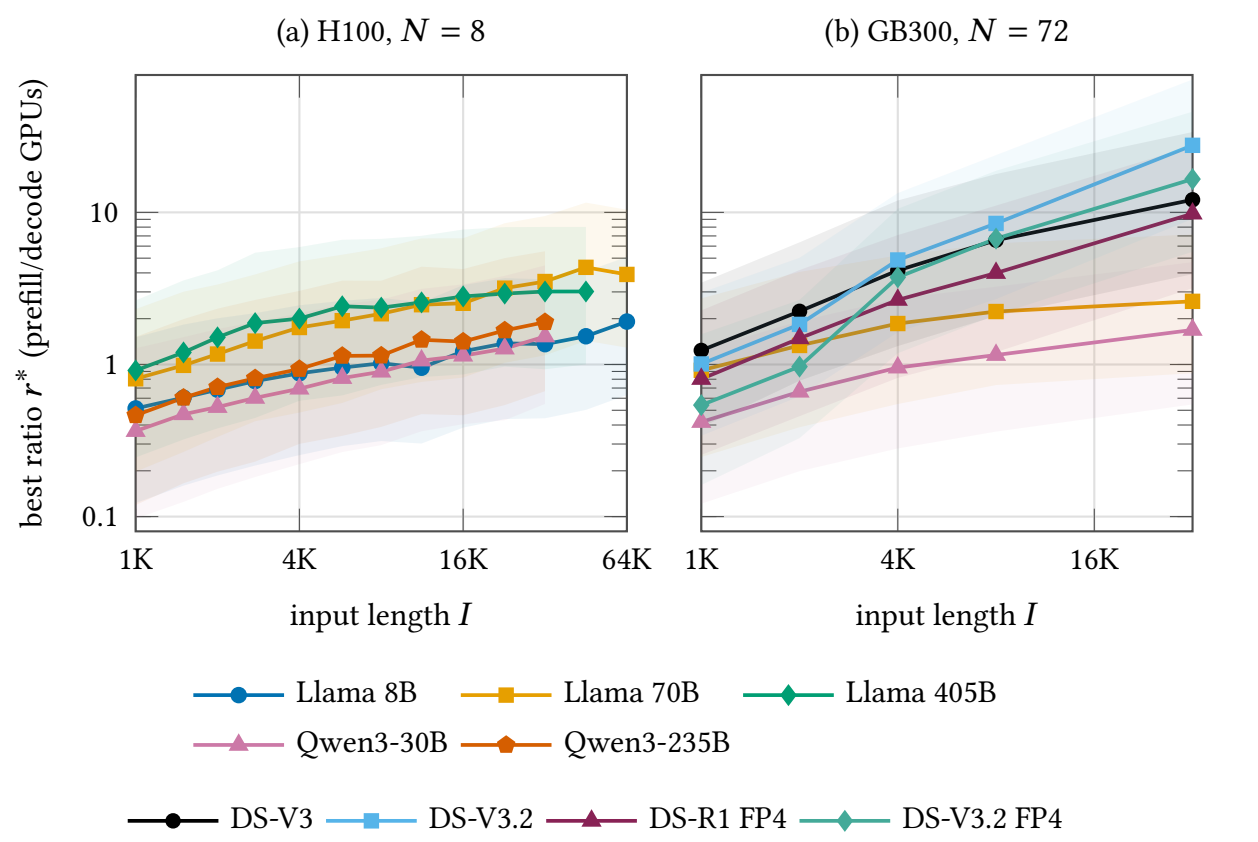

%% file: fig_hbm.tex
\begin{figure}[t]
\centering
\begin{tikzpicture}
\begin{axis}[osplit, width=0.96\columnwidth, height=4.8cm, xmode=log, log basis x=2, xtick={1024,4096,16384,65536}, xticklabels={1K,4K,16K,64K},
  xmin=1024, xmax=65536, ymin=0, ymax=100, xlabel={input length $I$ (tokens)}, ylabel={attention, \% of decode step time},
  legend style={at={(0.5,1.02)}, anchor=south}, legend columns=3]
\addplot[m8b] coordinates {(1024,65.49) (1536,68.46) (2048,70.86) (2816,75.34) (4096,77.73) (5888,80.34) (8192,82.84) (11520,83.86) (16384,87.51) (23296,86.77) (32768,88.84) (46336,86.66) (65536,90.18)};
\addlegendentry{Llama 8B}
\addplot[m70b] coordinates {(1024,32.75) (1536,34.12) (2048,39.38) (2816,42.71) (4096,49.47) (5888,51.86) (8192,54.94) (11520,59.77) (16384,58.37) (23296,66.27) (32768,71.8) (46336,68.42) (65536,64.7)};
\addlegendentry{Llama 70B}
\addplot[m405b] coordinates {(1024,7.195) (1536,10.14) (2048,9.741) (2816,11.85) (4096,15.3) (5888,13.35) (8192,20.21) (11520,21.18) (16384,14.55) (23296,14.65) (32768,15.62) (46336,15.56)};
\addlegendentry{Llama 405B}
\addplot[mq30] coordinates {(1024,48.3) (1536,53.8) (2048,55.86) (2816,59.18) (4096,61.76) (5888,65.6) (8192,66.67) (11520,68.85) (16384,69.25) (23296,74.21) (32768,71.12)};
\addlegendentry{Qwen3-30B}
\addplot[mq235] coordinates {(1024,18.49) (1536,21) (2048,22.76) (2816,24.13) (4096,30.28) (5888,29.35) (8192,37.06) (11520,31.18) (16384,38.29) (23296,32.09) (32768,33.02)};
\addlegendentry{Qwen3-235B}
\end{axis}
\end{tikzpicture}
\caption{Share of the decode step time spent in attention on H100, against input length.}
\vspace{-15pt}
\label{fig:hbm}
\end{figure}
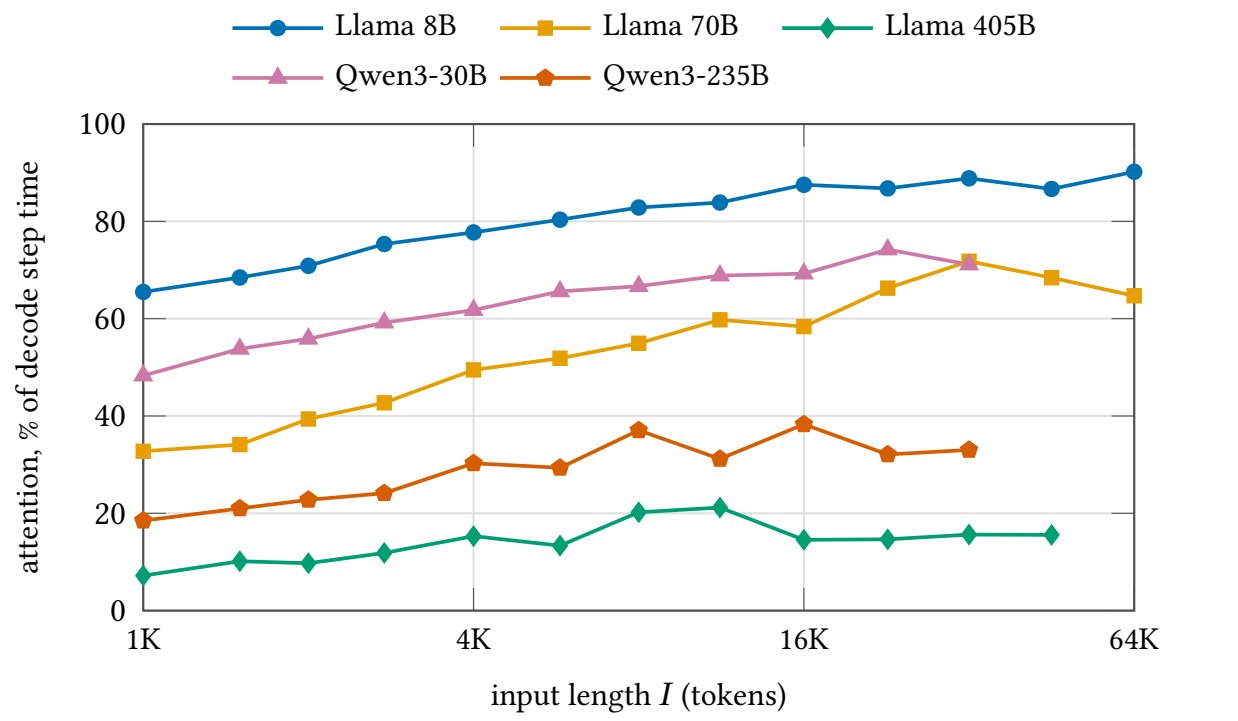